\documentclass[12pt]{article}
\usepackage{graphicx}
\usepackage{epsf,amsmath,bbold,amsfonts,stmaryrd}
\usepackage[utf8]{inputenc}
\usepackage[font=small,labelfont=bf,
   justification=justified, format=plain]{caption}
\usepackage{mathrsfs}
\usepackage{appendix}
\usepackage{amssymb}
\usepackage{float}
\usepackage{color} 
\usepackage[colorlinks]{hyperref}
\hypersetup{pageanchor=false,citecolor=blue,
  urlcolor=blue,linkcolor=blue}
\usepackage{indentfirst}
\usepackage{url}
\usepackage{float}
\usepackage{caption}
\usepackage{mathtools}
\usepackage{comment}
\usepackage[T1]{fontenc}
\usepackage{lmodern}
\usepackage[numbers,square,comma,sort&compress,merge]{natbib}
\usepackage{multibib}
\usepackage{enumerate} 
\usepackage{footnote}
\makesavenoteenv{tabular}
\makesavenoteenv{table}
\usepackage[normalem]{ulem}
\usepackage{multicol}
\usepackage{booktabs}
\usepackage{siunitx}
\usepackage{geometry}
\def\a{\alpha}

\renewcommand{\a}{\alpha}

\renewcommand{\d}{\delta}

\newcommand{\m}{\mu}
\newcommand{\n}{\nu}

\usepackage{amsmath}

\newcommand{\be}[1]{\begin{equation} \centering \label{#1}} 
\newcommand{\ee}{\end{equation}}
\def\bea{\begin{eqnarray}}
\def\eea{\end{eqnarray}}
\def\ba{\begin{array}}
\def\ea{\end{array}}
\def\bc{\begin{center}}
\def\ec{\end{center}}
\def\bl{\begin{flushleft}}
\def\el{\end{flushleft}}
\def\br{\begin{flushright}}
\def\er{\end{flushright}}
\def\bi{\begin{itemize}}
\def\ei{\end{itemize}}

\def\bt{\begin{tabular}}
\def\et{\end{tabular}}
\newtheorem{question}{Question}
\def\bq{\begin{question}}
\def\eq{\end{question}}
\newtheorem{definition}{Def}
\def\bd{\begin{definition}}
\def\ed{\end{definition}}
\newtheorem{answer}{Answer}
\def\ban{\begin{answer}}
\def\ean{\end{answer}}
\newtheorem{possibleanswer}{Possible answer}
\def\bpa{\begin{possibleanswer}\normalfont}
\def\epa{\end{possibleanswer}}
\newtheorem{theorem}{Theorem}
\def\bth{\begin{theorem}}
\def\eth{\end{theorem}}

\begin{document}
\vspace*{-5mm}
\begin{flushright}
HIP-2018-22/TH
\end{flushright}
\vspace{.3cm}
\thispagestyle{empty}

\begin{center}

{\bf \Large{Scale-invariant alternatives to general
    relativity. III. \\  
    The inflation--dark-energy connection}}\\
\vspace{.5cm}
{\textsc{Santiago Casas,$^{\dagger,}
    \footnote{\texttt{santiago.casas@cea.fr}}$
    ~~Georgios K. Karananas,$^{\ddagger,}
    \footnote{\texttt{georgios.karananas@physik.uni-muenchen.de}}$ 
    ~~Martin Pauly,$^{\ast,}
    \footnote{\texttt{m.pauly@thphys.uni-heidelberg.de}}$\\
    Javier Rubio$\,^{\ast,\S,}
    \footnote{\texttt{javier.rubio@helsinki.fi}}$ 
}}

\end{center}

\begin{center}

  \it{
    $^\dagger$\,AIM, CEA, CNRS, Université Paris-Saclay, 
Université Paris Diderot, \\
Sorbonne Paris Cité, F-91191 Gif-sur-Yvette, France

\vspace{.3cm}

$^\ddagger$\,Arnold Sommerfeld Center, 
Ludwig-Maximilians-Universit\"at M\"unchen,\\
Theresienstra{\ss}e 37, 80333 M\"unchen, Germany\\

\vspace{.3cm}

$^\ast$\,Institut f\"ur Theoretische Physik,
Ruprecht-Karls-Universit\"at Heidelberg, \\
Philosophenweg 16, 69120 Heidelberg, Germany\\

\vspace{.3cm}

$^\S$\,Department of Physics and Helsinki Institute of Physics, \\ 
PL 64, FI-00014 University of Helsinki, Finland

\vspace{.3cm}
}

\end{center}

\begin{abstract}

We discuss the cosmological phenomenology of biscalar--tensor models
displaying a maximally symmetric Einstein--frame kinetic sector and
constructed on the basis of scale symmetry and volume--preserving
diffeomorphisms.  These theories contain a single dimensionful
parameter $\Lambda_0$---associated with the invariance under the
aforementioned restricted coordinate transformations---and a massless
dilaton field. At large field values these scenarios lead to inflation
with no generation of isocurvature perturbations.  The corresponding
predictions depend only on two dimensionless parameters, which
characterize the curvature of the field--manifold and the leading
order behavior of the inflationary potential. For $\Lambda_0=0$ the
scale symmetry is unbroken and the dilaton admits only derivative
couplings to matter, evading all fifth force constraints. For
$\Lambda_0\neq 0$ the field acquires a run-away potential that can
support a dark energy dominated era at late times. We confront a
minimalistic realization of this appealing framework with observations
using a Markov-Chain-Monte-Carlo approach, with likelihoods from
present BAO, SNIa and CMB data. A Bayesian model comparison indicates
a preference for the considered model over $\Lambda$CDM, under certain
assumptions for the priors.  The impact of possible consistency
relations among the early and late Universe dynamics that can appear
within this setting is discussed with the use of correlation
matrices. The results indicate that a precise determination of the
inflationary observables and the dark energy equation--of--state could
significantly constrain the model parameters.

\end{abstract}

\newpage
\setcounter{page}{1}

\section{Introduction}
\label{sec:intro}

We have entered an era of precision cosmology. Cosmological parameters
are measured with unprecedented accuracy~\cite{Aghanim:2018eyx} and,
in addition to electromagnetic probes, gravitational wave
observations~\cite{Abbott:2016blz,Abbott:2016nmj,
TheLIGOScientific:2017qsa, Monitor:2017mdv} tightly constrain a
plethora of modified gravity
scenarios~\cite{Baker:2017hug,Creminelli:2017sry,Sakstein:2017xjx,
Ezquiaga:2017ekz, Langlois:2017dyl}.

In spite of the undeniable success of modern cosmology, the origin of
the present accelerated expansion of the Universe remains unknown. The
next decade of observations---with an upcoming generation of galaxy
redshift surveys such as
Euclid~\cite{Amendola:2016saw,Laureijs:2011gra} or LSST
\cite{Mandelbaum:2018ouv}---will be of crucial importance for
determining whether this phase arises due to an inert cosmological
constant or rather a dynamical dark energy component. The combination
of these surveys with Stage--IV Cosmic Microwave Background (CMB)
observations~\cite{Abazajian:2016yjj} will also pin down the
inflationary parameters, setting the stage for more fundamental
questions on the relation between the early and late Universe. Indeed,
although inflation and dark energy are usually treated as two
independent epochs, they might be closely related, as happens for
instance in quintessential inflationary
models~\cite{Peebles:1998qn,Spokoiny:1993kt,Brax:2005uf,
Dimopoulos:2017zvq,Rubio:2017gty,Akrami:2017cir,
Garcia-Garcia:2018hlc} or in certain theories invariant under
dilatations~\cite{GarciaBellido:2011de,Trashorras:2016azl,Casas:2017wjh}. 
A potential confirmation of this appealing hypothesis might completely
change our understanding of modern cosmology.

In the last few years there has been renewed interest in the
implications of scale and conformal symmetries and many of their
aspects---both formal and phenomenological---have been thoroughly
investigated~\cite{Englert:1976ep,Wetterich:1987fm,Wetterich:1987fk,
Wetterich:1994bg,Bardeen:1995kv,Meissner:2006zh,Shaposhnikov:2008xb,
Shaposhnikov:2008xi,
GarciaBellido:2011de,Blas:2011ac,GarciaBellido:2012zu,Bezrukov:2012hx,
Monin:2013gea,Armillis:2013wya,Tavares:2013dga,Gretsch:2013ooa,
Khoze:2013uia,Rubio:2014wta,Karam:2015jta,
Karananas:2015eha,Karananas:2015ioa,
DiVecchia:2015jaq,Lucat:2016eze,Trashorras:2016azl, Karananas:2016grc,
Ferreira:2016vsc,Bianchi:2016viy,
Karananas:2016kyt,Karam:2016rsz,Ferreira:2016wem,Ferreira:2016kxi,
Shkerin:2016ssc,Karananas:2017mxm,Rubio:2017gty,
DiVecchia:2017uqn,Guerrieri:2017ujb,Tokareva:2017nng,Gillioz:2017ooc,
Karananas:2017zrg,Lucat:2017wtu,Casas:2017wjh,
Ferreira:2018itt,Ferreira:2018qss,Shaposhnikov:2018xkv,Lucat:2018slu,
Shaposhnikov:2018jag,Burrage:2018dvt,SravanKumar:2018tgk,Lalak:2018bow,
Gorbunov:2018llf,Barnaveli:2018dxo,Quiros:2018ryt}.  In this paper, we
focus on the cosmological consequences of a general class of
biscalar--tensor models first introduced in Ref.~\cite{Blas:2011ac}
which are invariant under volume--preserving
diffeomorphisms\,\footnote{The precise definition of these
transformations can be found in the next section.} and display
spontaneous breaking of scale symmetry.  We restrict ourselves to
theories that contain at most two derivatives of the fields, such that
the particle spectrum comprises only healthy degrees of freedom. In
addition, we are interested only in models whose Einstein--frame
target manifold is maximally symmetric during inflation and, more
precisely, globally hyperbolic
\cite{Karananas:2016kyt,Klein:2017hch}.\footnote{The reasons for this
choice will become apparent later, but the essence is that under this
assumption the arbitrariness in the construction of the corresponding
action is greatly reduced.} Moreover, we require the equations of
motion governing the dynamics of the theories under consideration to
admit Minkowski, de Sitter and anti de Sitter vacuum solutions, since
these might be essential for the eventual quantization of the theory.
Finally, when needed, we assume the existence of a hierarchy between
the inflationary and particle--physics scales, similar to that between
the Planck and electroweak scales.

Even though we will go into further details in what follows, let us
spell out some of the most intriguing features of these specific
models. On general grounds, these theories contain a single
dimensionful parameter $\Lambda_0$ associated with the invariance of
the action under volume-preserving coordinate transformations. For
$\Lambda_0=0$, one of the scalar fields, which we will
call~\emph{dilaton}, becomes the Goldstone boson of the spontaneously
broken scale symmetry. The combination of gravity and dilatation
invariance forces this field to have only derivative couplings to
matter. Consequently, the fifth--force effects associated with the
dilaton are highly suppressed in this particular
context~\cite{Shaposhnikov:2008xb,GarciaBellido:2011de,Ferreira:2016kxi,
Burrage:2018dvt}. Additionally, scale invariance forbids the
generation of isocurvature perturbations during the inflationary stage
due to the presence of a conserved (scale) current that effectively
reduces the biscalar theory to a single--field
scenario. Interestingly, these theories admit also an
``$\a$--attractor''
solution~\cite{Ferrara:2013rsa,Kallosh:2013yoa,Galante:2014ifa} for
the spectral tilt and tensor--to--scalar ratio
\cite{Karananas:2016kyt}. For $\Lambda_0\neq 0$, the dilatation
symmetry is explicitly broken. The combination of this specific
symmetry--breaking term with the omnipresent nonminimal coupling to
gravity of scalar--tensor theories leads to a~\emph{unique}
quintessential potential for the dilaton
field~\cite{Shaposhnikov:2008xb}. For sufficiently small values of
$\Lambda_0$, all the inflationary properties mentioned previously are
approximately realized and the dilaton remains an almost massless
degree of freedom potentially responsible for the current accelerated
expansion of the Universe. This, in turn, can lead to a set of
nontrivial consistency conditions between the inflationary observables
and the dark energy equation--of--state parameter, which could be
tested with future cosmological observations.

The paper is organized as follows. In
Sec.~\ref{sec:scal_inv_two_field}, we introduce the notion of
transverse diffeomorphisms and, closely following
Ref.~\cite{Blas:2011ac}, we construct the most general class of
scale--invariant biscalar models invariant under this type of
transformations. In Sec.~\ref{sec:EF}, we recast the obtained set of
models in the Einstein--frame, where the gravitational part of the
action takes the usual Einstein--Hilbert form. After discussing the
general features above, we focus on models involving a
maximally--symmetric field manifold in the Einstein frame. The
cosmological consequences of this broad class of theories are
considered in Sec.~\ref{sec:inf_DE}, while in Sec.~\ref{sec:confront},
we make use of a Markov Chain Monte Carlo approach to confront a
particular realization of our scenario with present data sets and
discuss the chances of differentiating it from other cosmological
scenarios such as $\Lambda$CDM. Finally, our conclusions are presented
in Sec.~\ref{sec:concl}.

\section{Scale-invariant biscalar models}
\label{sec:scal_inv_two_field}

Our current understanding of the gravitational interaction is based on
a massless spin--two field, the graviton. In general relativity, this
degree of freedom is associated with general coordinate
transformations or diffeomorphisms (Diffs). At the infinitesimal
level, these transformations take the form
\begin{equation}
x^\mu\to x^\mu (x)+\d x^\mu(x)\ ,
\end{equation}
with $\d x^\mu$ arbitrary. In spite of this ``traditional''
association, the minimal group leading to graviton
excitations~\emph{is not} the group of general coordinate
transformations, but rather the subgroup spanned by the transverse
vectors
\begin{equation}
\label{eq:tdiff_vecs}
\d x^\mu =\xi^\mu\ ,~~~\text{with}~~~\partial_\mu \xi^\mu=0\ .
\end{equation}
In what follows we will refer to these transformations as
volume--preserving, restricted, or transverse diffeomorphisms (TDiff),
interchangeably. It should be clearly stated that, in general,
theories invariant under~(\ref{eq:tdiff_vecs}) propagate an extra
scalar degree of freedom related to the metric determinant on top of
the two graviton polarizations.

Contrary to what happens in diffeomorphism--invariant theories, the
requirement of invariance under TDiffs~(\ref{eq:tdiff_vecs}) does not
completely determine the form of the action. In particular, it is
always possible to include arbitrary functions of the metric
determinant $g\equiv-\text{det}(g_{\m\n})$ in the Lagrangian density,
since this quantity transforms as a scalar under volume--preserving
diffeomorphisms. As shown in Ref.~\cite{Blas:2011ac}, the most general
TDiff action that is also invariant under scale transformations
[$g_{\m\n}(x)$ is the metric, $\phi(x)$ a scalar field with scaling
dimension one]
\begin{align}
\label{stsupp}
g_{\m\n}(x)\mapsto g_{\m\n}(\lambda\, x) \ ,\hspace{10mm}
\phi(x)\mapsto \lambda\,\phi(\lambda\, x)\ ,\hspace{10mm}
\lambda=\text{const.} \ ,
\end{align}
takes the form
\begin{equation}
\begin{aligned}
\label{Tdiffinit}
\mathcal S=\int d^4x \sqrt{g}\Bigg\{ \frac{\phi^2f(g)}{2} R&-
\frac{\phi^2}{2}\Big[G_{1}(g) \left(\partial g\right)^2-2 \,G_{2}(g)
\left(\partial g\right)\left(\phi^{-1}\partial\phi \right)\\
&+G_{3}(g) \left(\phi^{-1}\partial \phi \right)^2 \Big] -
\phi^4v(g)\Bigg\}\ ,
\end{aligned}
\end{equation}
with $f$, $G_1$, $G_2$, $G_3$ and $v$ arbitrary functions of the
metric determinant. For general choices of these
\emph{theory--defining functions}, the action~(\ref{Tdiffinit})
contains three propagating degrees of freedom on top of the scalar
field $\phi$: the two graviton polarizations and a new scalar
associated with the metric determinant.\footnote{The additional degree
of freedom is only absent for very particular choices of the
theory-defining functions, leading either to general relativity or to
unimodular
gravity~\cite{vanderBij:1981ym,Wilczek:1983as,Buchmuller:1988wx,
Unruh:1988in,
Weinberg:1988cp,Henneaux:1989zc,Buchmuller:1988yn}. Interestingly,
these two limiting cases are completely equivalent at the classical
level but not necessarily when quantum corrections are taken into
account \cite{Alvarez:2015pla,Alvarez:2015sba}.}
The existence of this additional degree of freedom can be made
explicit by rewriting the above action in a Diff-invariant form. To
this end, we first transform~(\ref{Tdiffinit}) to an arbitrary
coordinate frame (i.e we perform a \textit{general} coordinate
transformation with Jacobian $J(x)\neq 1$), to
obtain~\cite{Blas:2011ac}
\begin{equation}
\begin{aligned}
\label{eq:lagr_ga}
\mathcal S= \int d^4x \sqrt{g}\Bigg\{
\frac{\phi^2f\left(\frac{g}{a}\right) }{2}
R&-\frac{\phi^2}{2}\Big[G_{1}\left(\frac{g}{a}\right) \left(\partial
g/a\right)^2-2 \,G_{2}\left(\frac{g}{a}\right) \left(\partial
g/a\right)\left(\phi^{-1}\partial\phi \right)\\
&+G_{3}\left(\frac{g}{a}\right) \left(\phi^{-1}\partial \phi \right)^2
\Big]
-\phi^4v\left(\frac{g}{a}\right)-\frac{\Lambda_0}{\sqrt{g/a}}\Bigg\} \
,
\end{aligned}
\end{equation}
with $a(x)\equiv J(x)^{-2}$ and $\Lambda_0$ a unique scale
symmetry--breaking term that arises as an integration constant in the
original TDiff formulation.\footnote{ For more details on this point,
the interested reader is referred to Ref.~\cite{Blas:2011ac}.}
Promoting $a(x)$ to a (dynamical) compensator field transforming under
Diffs as
\begin{equation}
\delta_\xi a =\xi^\mu \partial_\mu a +2 a \partial_\mu \xi^\mu\ ,  
\end{equation}
the Lagrangian density in (\ref{eq:lagr_ga}) can equivalently be
written as
\begin{equation}
\begin{aligned}
\label{eqn:hdm_lagrangianL} 
\frac{\mathscr L}{\sqrt{g}}= \frac{\phi^2f(\tilde\theta) }{2}
R&-\frac{\phi^2}{2}\Big[G_{1}(\tilde\theta) (\partial
\tilde\theta)^2+2 \,G_{2}(\tilde\theta) (\partial
\tilde\theta)\left(\phi^{-1}\partial\phi \right)+G_{3}(\tilde\theta)
\left(\phi^{-1}\partial \phi\right)^2 \Big] \\ &-\phi^4v(\tilde\theta)
-\frac{\Lambda_0}{\sqrt{\tilde\theta}}\ ,
\end{aligned}
\end{equation}
with $\tilde\theta\equiv g/a>0$~\cite{Blas:2011ac,Karananas:2016grc}.
This expression is, by construction, invariant under general
coordinate transformations and reduces to the TDiff form
\eqref{Tdiffinit} in the $a=1$ gauge. Given the (classical)
equivalence of the TDiff-- and Diff--invariant
formulations~\cite{Blas:2011ac,Shaposhnikov:2008xb}, we will work in
what follows with the more familiar diffeomorphism--invariant
form. Note also that small choices of $\Lambda_0$ are technically
natural \cite{tHooft:1979rat} (see also
Refs.~\cite{Alvarez:2015pla,Alvarez:2015sba}). Indeed, for
$\Lambda_0=0$, the action associated with~\eqref{eqn:hdm_lagrangianL}
is invariant under scale transformations, which are now internal. This
means that the coordinates are kept fixed, while the various fields
change as\,\footnote{This should be compared with the
transformation~(\ref{stsupp}).}
\begin{align}
\label{stsupp2}
 g_{\m\n}(x)\mapsto \lambda^2 g_{\m\n}(x),\hspace{10mm} \phi(x)
\mapsto \lambda\,\phi(x),\hspace{10mm} \tilde\theta(x)\mapsto
\tilde\theta(x) \ .
\end{align}

\section{Einstein-frame formulation}
\label{sec:EF}

The phenomenological consequences of the theories under consideration
are most easily studied in the \emph{Einstein frame}, in which the
gravitational part of the action takes a ``canonical form.'' Requiring
the existence of a well-defined graviton propagator at all field
values, i.e. $\phi^2f(\tilde\theta) > 0$, we can perform the Weyl
rescaling $g_{\mu\nu}\to M_P^2/(\phi^2f(\tilde\theta))\, g_{\mu\nu}$,
to rewrite
\eqref{eqn:hdm_lagrangianL} as
\bea
\label{act_max_sym}
\frac{\mathscr L}{\sqrt{g}}=\frac{M_P^2}{2}R &-&\frac{1}{2} \Big[M_P^2
K_1(\tilde\theta) (\partial \tilde\theta)^2+ 2M_PK_2(\tilde\theta)
(\partial \tilde\theta)(\partial \tilde\Phi)+K_3(\theta) (\partial
\tilde\Phi)^2 \Big] \nonumber\\
&-& \, U(\tilde\theta) -
\frac{\Lambda_0}{f^2(\tilde\theta)\sqrt{\tilde\theta}} e^{-4
\tilde\Phi/M_P}\ ,
\eea
with $M_P=2.48\times 10^{18}$ GeV the reduced Planck mass.  In the
above expression, we introduced the following
$\tilde\theta$--dependent functions
\begin{eqnarray}\label{Ks0}
K_1(\tilde\theta)&\equiv&\dfrac{G_1(\tilde\theta)}{f(\tilde\theta)}+
\dfrac{3}{2}\left(\dfrac{f'(\tilde\theta)}{f(\tilde\theta)}\right)^2\
,\hspace{5mm}
K_2(\tilde\theta)\equiv\dfrac{G_2(\tilde\theta)}{f(\tilde\theta)}+3\,
\dfrac{f'(\tilde\theta)}{f(\tilde\theta)}\ , \\
K_3(\tilde\theta)&\equiv&
6+\dfrac{G_3(\tilde\theta)}{f(\tilde\theta)}\ , \hspace{30mm}
U(\tilde\theta)=\frac{M_P^4 v(\tilde\theta)}{f^2(\tilde\theta)}  \label{Ks0b}\ ,
\end{eqnarray}
and used primes to denote derivatives with respect to
$\tilde\theta$. Note that the rescaled field
\be{eq:phi_lower_phi_cap}
\tilde\Phi\equiv M_P\ln\left(\frac{\phi}{M_P}\right) \ ,
\ee 
is defined in such a way that the scale
transformations~(\ref{stsupp2}) act on it as a shift.

The non-diagonal kinetic terms in \eqref{act_max_sym} can be 
diagonalized by considering an additional field
redefinition~\cite{Blas:2011ac,Karananas:2016kyt}
\begin{align}\label{Phishift}
\tilde\Phi \to \Phi= \tilde\Phi - M_P \int^{\tilde\theta}
d\tilde\theta'\, \frac{K_{2}(\tilde\theta')}
{K_{3}(\tilde\theta')}\ .
\end{align}
Once this is performed, we obtain the Lagrangian density
\be{act_max_sym_2}
\frac{\mathscr L}{\sqrt{g}}=\frac{M_P^2}{2}R -\frac{1}{2}\Bigg[M_P^2
K(\tilde\theta)(\partial \tilde\theta)^2+K_3(\tilde\theta) (\partial
\Phi)^2 \Bigg]-U(\tilde\theta)-U_{\Lambda_0}(\tilde\theta,\Phi)\ ,
\ee
with
\be{eq:pot_lambda}
U_{\Lambda_0}(\tilde\theta,\Phi)=\Lambda_0\, K_\Lambda(\tilde \theta)
\,e^{-4 \Phi/M_P } \ ,
\ee
and
\be{VL}
K(\tilde\theta)=\frac{K_1(\tilde\theta)
K_3(\tilde\theta)-K_2^2(\tilde\theta)}{K_3(\tilde\theta)}\ ,
\hspace{15mm} K_\Lambda=
\frac{1}{f^2(\tilde\theta)\sqrt{\tilde\theta}}\exp \left({4
\int\,\frac{K_2(\tilde\theta')}{K_3(\tilde\theta') }}\, d
\tilde\theta' \right) \ .
\ee 
In order to ensure the absence of ghost-like excitations and to
prevent the potential appearance of anti-de Sitter regimes, we will
require the $\tilde\theta$-dependent functions in these expressions to
be positive at all field values, namely\,\footnote{The
shift~(\ref{Phishift}) excludes the $K_3(\tilde\theta)=0$
case. The choice $K(\tilde\theta)=0$ is also excluded,
since in such a case the target manifold would be
one--dimensional. This means that one of the two propagating degrees
of freedom becomes nondynamical.}  
 \begin{equation}
K(\tilde\theta) > 0\ ,\hspace{10mm} K_3(\tilde\theta) > 0\,,
\hspace{10mm}  U(\tilde\theta)\geq 0\ ,\hspace{10mm} \Lambda_0\,
K_\Lambda(\tilde\theta)\geq 0\,.
\end{equation}
Note that these conditions do not restrict the derivatives of the
corresponding functions, which could be negative for particular field
ranges, allowing for instance for limited tachyonic instabilities.

For $\Lambda_0=0$, the Lagrangian density
\eqref{act_max_sym_2} acquires an emergent shift symmetry $\Phi\to\Phi
+M_P\,C$ with $C$ a constant. This symmetry is nothing else than a
manifestation of the non-linear realization of the original scale
symmetry~(\ref{stsupp}) (or equivalently~(\ref{stsupp2})) that the
theory exhibits in the scaling frame~(\ref{eqn:hdm_lagrangianL}). The
field $\Phi$ is therefore identified as the Goldstone boson or
\emph{dilaton} associated with the spontaneous breaking of scale
invariance. For $\Lambda_0\, K_\Lambda (\tilde\theta)> 0$, the
symmetry is explicitly broken and the dilaton acquires the runaway
potential~(\ref{eq:pot_lambda}).

Given the Lagrangian density in the form~(\ref{act_max_sym_2}), it is
still possible to perform additional field redefinitions to modify the
precise structure of the theory-defining functions $K(\tilde\theta)$,
$K_3(\tilde\theta)$, etc. For instance, if $K(\tilde\theta)\neq 0$, we
can introduce a variable\,\footnote{Here, $\tilde\theta_0$ is an
arbitrary integration constant ensuring that
$\theta(\tilde\theta_0)=0$.}
\begin{align}
\theta=\int_{\tilde\theta_0}^{\tilde\theta}
d\tilde\theta'\sqrt{\left|\frac{K_1(\tilde\theta') K_3(\tilde\theta') 
-K_2^2(\tilde\theta')}{K_3(\tilde\theta')}\right|}\ ,
\end{align}
in terms of which the kinetic term of $\tilde\theta$ becomes
canonical.\footnote{In general, the kinetic sector can always be
diagonalized. However, the kinetic mixing among the fields
\emph{cannot} be removed, unless the target manifold is flat.} In this
case, we get the following Lagrangian density
\be{act_max_sym_3}
\frac{\mathscr L}{\sqrt{g}}=\frac{M_P^2}{2}R -\frac{1}{2} \Bigg[M_P^2  
(\partial \theta)^2+K_3( \theta)(\partial
\Phi)^2 \Bigg]-U( \theta)-U_{\Lambda_0}(\theta,\Phi)\ .
\ee
This freedom to perform field redefinitions can be trivially
understood once the scalars are viewed as the coordinates of the
two-dimensional field manifold.  In fact, this interpretation allows
to rewrite the Einstein-frame Lagrangian in the explicitly covariant
form
\begin{equation}\label{lagrangianE} 
\frac{\mathcal{L}}{\sqrt{g}}=\frac{M_P^2}{2}R-
\frac{1}{2} \gamma_{ab} g^{\mu\nu}\partial_{\mu}\varphi^a
\partial_{\nu}\varphi^b- V(\varphi)\ .
\end{equation}
Here, the latin indices $a,b,...=1,2$ denote the two real scalars
present in the model, $\gamma_{ab}$ is the metric in this field space
and
\be{eq:pot_collective}
V(\varphi)=U(\varphi)+U_{\Lambda_0}(\varphi) \ .
\ee
The variation of the action associated with the Lagrangian
density~(\ref{lagrangianE}), leads to the Einstein and Klein-Gordon
equations, respectively

\begin{equation}\label{ein}
M_P^2\,G_{\mu\nu}=-\gamma_{ab}\left(\partial_\mu\varphi^a\partial_\nu\varphi^b  
-\frac{1}{2} g_{\mu\nu} g^{\rho\sigma}
\partial_\rho\varphi^a\partial_\sigma\varphi^b\right) +V g_{\mu\nu}\ ,
\end{equation} 
\begin{equation}\label{kg}
{\Box}\varphi^c+ g^{\mu\nu}\mathscr G^c_{ab}
\partial_\mu\varphi^a\partial_\nu\varphi^b=\gamma^{cd} V_{,d}\ ,
\end{equation} 
where $G_{\mu\nu}$ is the Einstein tensor computed from the
Einstein--frame~\emph{space-time metric} $g_{\mu\nu}$ and $\mathscr
G^c_{ab}$ is the (symmetric) affine connection computed from the
Einstein--frame~\emph{field-space metric} $\gamma_{ab}$,
i.e.\footnote{As customary, the comma denotes partial derivative.}
\be{eq:christ_field_space}
\mathscr
G^c_{ab}=\frac{1}{2}\gamma^{cd}\left(\gamma_{da,b}+\gamma_{db,a} 
-\gamma_{ab,d}\right) \ .
\ee
 \vspace{1mm}

\subsection{Scale current and single-field dynamics}

In the absence of the dimensionful parameter $\Lambda_0$, the scale
invariance of the theories under consideration leads to the existence
of a (covariantly) conserved current, which can be obtained from
Noether's theorem. In the Einstein frame, it reads
\be{eq:cur_ein_fram_0} 
J^\m = - \gamma_{ab}\partial^\m\varphi^a\Delta\varphi^b \ ,
\ee
with $\Delta\varphi^a$ denoting the infinitesimal action of
dilatations on the fields. Note that both the explicit form of $\Delta
\varphi^a$ and the current depend on the variables under
consideration.  For instance, for the variables in
Eq.~(\ref{act_max_sym}) we have $\Delta\varphi^a\equiv(\Delta
\tilde\theta,\Delta\tilde\Phi)=(0,M_P)$, and
\be{eq:cur_ein_fram_1} 
J^\m = -M_P
\left(K_3(\tilde\theta)\partial^\m\tilde\Phi + M_P
  K_2(\tilde\theta)\partial^\m \tilde\theta\right)\ .
\ee
For the ones in Eq.~(\ref{act_max_sym_2}), we see that the
infinitesimal transformation corresponds to $\Delta\varphi^a=(\Delta
\theta,\Delta\Phi)=(0,M_P)$, and the current is given by
\be{eq:cur_ein_fram_2}
J^\m = -  M_P
K_3(\theta) \partial^\m \Phi \ .
\ee
From either Eq.~(\ref{eq:cur_ein_fram_1})
or~(\ref{eq:cur_ein_fram_2}), we find that the (covariant) divergence
of the scale current takes the form
\begin{equation}\label{eq:currentcov}
\frac{1}{\sqrt{g}}\partial_\mu\left(\sqrt{g}J^\mu\right)
=4U_{\Lambda_0} \ ,
\end{equation}
clearly showing that the above indeed vanishes only for $\Lambda_0=0$.
For homogeneous fields in the cosmologically relevant
Friedmann-Lemaitre-Robertson-Walker background, this equation takes
the very simple form,
\begin{equation}\label{cchom}
\frac{1}{a^3}\frac{d}{d t}\left(a^3
\gamma_{ab}\dot\varphi^a\Delta\varphi^b\right)= 4 U_{\Lambda_0}\ , 
\end{equation}
with $a=a(t)$ the scale factor and the dots standing for derivatives
with respect to the coordinate time $t$. For small $\Lambda_0$ (and/or
sufficiently large dilaton expectation values), the contribution of
the symmetry breaking term in the right-hand side of this equation can
be safely neglected. In this limit, the quantity $a^3
\gamma_{ab}\dot\varphi^a\Delta\varphi^b$ becomes approximately
conserved, such that $\gamma_{ab}\dot\varphi^a\Delta\varphi^b$
approaches zero as the Universe expands. For the particular set of
variables in Eq.~(\ref{act_max_sym_2}), this statement takes the
intuitive form
\be{eq:phi_frozen}
\frac{d \Phi}{d N} \propto\frac{1}{H K_3(\theta)}e^{-3N} \ ,
\ee
with $H$ the Hubble parameter and $N$ the number of e-folds. An
immediate consequence of this equation is that $d\Phi/dN=0$ is
actually an attractor solution, leading to an effective constraint in
the $\lbrace h,\chi\rbrace$ plane \cite{GarciaBellido:2011de}.  The
existence of this attractor is of course a physical statement
independent of the frame in which the scale current is computed. In
particular, one could perform the same computation in the scaling
frame~(\ref{eqn:hdm_lagrangianL}). In this case, it is simpler to
obtain the precise expression for the current from Noether's theorem,
\be{eq:noeth_cur_def}
 J^\m = \frac{\d \mathscr L}{\d (\partial_\m g_{\n\lambda})}\Delta 
g_{\n\lambda} +\frac{\d \mathscr L}{\d (\partial_\m
  \phi)}\Delta\phi^i \,.
\ee 
Taking into account the infinitesimal form of~(\ref{stsupp2}), namely
$\Delta g_{\m\n} = -2 g_{\m\n}$ and $\Delta\phi=\phi$, we get
\be{eq:cur_jord_fram}
J^\m = -\frac{1}{2}\left[
\left(G_3(\tilde\theta)+6f(\tilde\theta)\right)\partial^\m \phi^2+
2\phi^2 \left(G_2(\tilde\theta)+3f'(\tilde\theta)\right)\partial^\m
\tilde\theta \right] \ .
\ee
This expression is nothing else than the conformally--transformed
version of the Einstein frame current~(\ref{eq:cur_ein_fram_1}), as
can be easily verified by taking into account the Weyl rescaling of
the metric together with the
relations~(\ref{Ks0})~--~(\ref{Phishift}).

\section{Inflation and dark energy in a single shot}
\label{sec:inf_DE}

The kinetic sector of~(\ref{act_max_sym_2}) constitutes a
nonlinear sigma model.  The associated (Gauss) curvature of the
Einstein--frame target manifold in Planck units is given by
\be{gauss}
\kappa(\tilde \theta)=\frac{K_3'(\tilde \theta)F'(\tilde
\theta)-2F(\tilde\theta) K''_3(\tilde\theta)}{4F^2(\tilde\theta)} \ ,
\ee
with $F(\tilde\theta)\equiv K(\tilde\theta)K_3 (\tilde\theta)$.  It
should be obvious at this point that without specifying the various
theory-defining functions, it is not possible to extract any detailed
information about the dynamics of the theory. However, for
inflationary models in which $\kappa(\tilde \theta)$ is
constant---corresponding to a maximally symmetric target
manifold---the situation simplifies considerably. The reason is that
in that case the above equation can be straightforwardly integrated to
obtain~\cite{Karananas:2016kyt}
\be{req_gho_free_2} 
K(\tilde\theta)=-\frac{ K^{'2}_3(\tilde\theta)}{4 \, K_3(\tilde\theta)
( \kappa K_3(\tilde \theta)+c)}\ ,
\ee
with $c$ an arbitrary constant. Assuming that both $U$ and $K_\Lambda$ 
are analytic functions of $\tilde \theta$ (such that they can be
expressed in term of $K_3$), we can rewrite \eqref{act_max_sym_2} as
\begin{equation}\label{MSattractor}
\frac{\mathscr L}{\sqrt{g}}=\frac{M_P^2}{2}R -
\frac{1}{2}\Bigg[-\frac{M_P^2(\partial \Theta)^{2}}{4\, \Theta (\kappa
\Theta+c)}+ \Theta(\partial \Phi)^2 \Bigg] - U(\Theta)-\Lambda_0\,
K_\Lambda(\Theta)\,e^{-4 \Phi/M_P }\ ,
\end{equation}
where we have defined a variable $\Theta\equiv K_3(\theta)$ to stress
the fact that the function $K_3$ \textit{itself} plays the role of
a~\emph{dynamical degree of freedom}.  The requirement that both
fields have healthy kinetic terms imposes the restrictions
\be{}
\kappa \Theta+c <0 \ ,\hspace{25mm}\Theta>0 \ .
\ee
Maximally--symmetric scale--invariant models can naturally support
inflation, while providing a unique dark--energy dominated era. To
understand this, let us focus on the pole structure
of~(\ref{MSattractor}). The kinetic term for the $\Theta$ field in
this expression contains two poles, located at $\Theta=0$ and
$\Theta=-c/\kappa$, respectively.  The presence of these poles
translates into an effective stretching of the canonically normalized
field $\theta$, namely\,\footnote{For a detailed discussion on the
connection between this approach and the existence of stationary
points along the inflationary trajectory see
Refs.~\cite{Rubio:2018ogq,Artymowski:2016pjz}.}
\begin{equation}\label{eq:cantheta}
\theta=\int^\theta \frac{d\Theta}{\sqrt{-4\,  \Theta
(\kappa \Theta+c)}} \hspace{5mm}\rightarrow \hspace{5mm} 
\Theta =
\begin{cases}
\exp\left(-2\sqrt{-\kappa}\,\theta\right)\,,\hspace{5mm}{\rm
for}\hspace{5mm} c=0 \ , \\ \frac{c}{-\kappa}\cosh
(\sqrt{-\kappa}\,\theta) \,, \hspace{5mm}{\rm for}\hspace{5mm} c\neq 
0\ .
\end{cases}
\end{equation}
For $c=0$, the two poles coincide and the stretching in $\theta$ is
exponential, with $\Theta=0$ corresponding to $\theta=\infty$. For
$c\neq 0$, the stretching of $\theta$ is restricted to a compact field
range around $\theta=0$.

\begin{figure}
{\begin{center}
\includegraphics[scale=0.5]{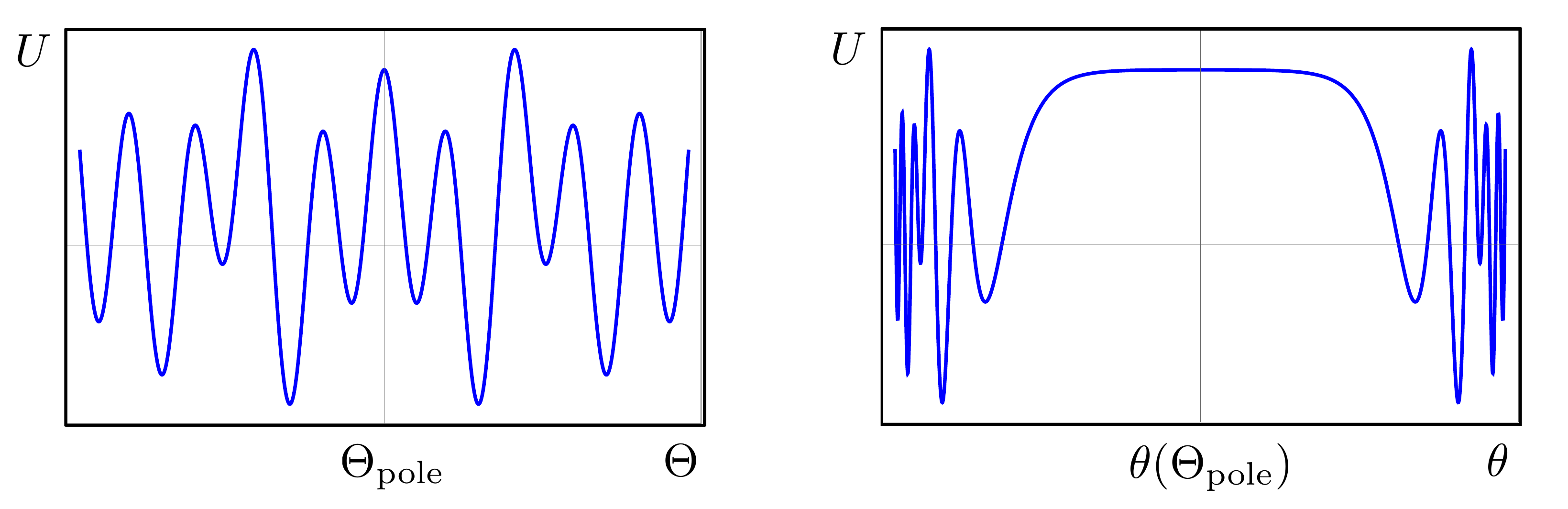}
\end{center}}
\caption{The effect of the Einstein-frame kinetic pole structure in
\eqref{MSattractor} for a generic potential $U(\Theta)$. The presence
of a pole at a value $\Theta_{\rm pole}$ translates into an effective
stretching of the canonically normalized field $\theta$ in
\eqref{eq:cantheta} and the associated flattening of the potential
around $\theta(\Theta_{\rm pole})$. This allows for inflation with the
usual slow--roll conditions even if the original potential was not
sufficiently flat.}
\label{fig:stretch}  
\end{figure}

This flattening of the potential for the canonically normalized field
$\theta$ allows for inflation with the usual slow-roll
conditions~\cite{Galante:2014ifa,Karananas:2016kyt,Rubio:2018ogq},
cf. Fig~\ref{fig:stretch}. For sufficiently small values of
$\Lambda_0$ (and/or sufficiently large values of the dilaton field
$\Phi$), the contribution of the $U_{\Lambda_0}$ term
in~\eqref{MSattractor} is subdominant and can be safely neglected. In
the absence of this symmetry--breaking term, the conservation of the
dilatation current~(\ref{eq:cur_ein_fram_0}) leads to the attractor
behavior~(\ref{eq:phi_frozen}) and forces the dilaton to freeze at a
given value, say $\Phi_0$, during the whole inflationary evolution. As
first proved in Ref.~\cite{GarciaBellido:2011de}, this reduces the
number of dynamical variables by one such and avoids the generation of
dangerous isocurvature fluctuations (see also
Ref.~\cite{Ferreira:2018qss}).

For potentials allowing a graceful inflationary exit, the inflaton
field $\Theta$ will undergo damped oscillations after the end of
inflation and will eventually relax to the ground state of $U(\Theta)$
via particle production. Although the shape of the potential in this
transition phase is~\emph{a priori} arbitrary, its precise low--energy
form can be restricted on phenomenological grounds. To see this, let
us neglect for the time being the term proportional to $\Lambda_0$ in
Eq.~\eqref{MSattractor} and consider the existence of stable solutions
involving constant field values $\Phi=\Phi_0$ and
$\Theta=\Theta_0$. Demanding $U'(\Theta_0)=0$, we obtain
[cf.~\eqref{Ks0b}]
\begin{equation}
f(\Theta_0)v'(\Theta_0)-2f'(\Theta_0)v(\Theta_0)=0\,.
\end{equation}
The Ricci scalar associated with this field configuration can be
easily determined by tracing the Einstein equation~(\ref{ein}) over
spacetime indexes. Taking into account that the contributions from the
field derivatives in this expression vanish for constant field values
together with the second relation in \eqref{Ks0b}, we obtain
\begin{equation} R=-4M_P^2\frac {v(\Theta_0)} {f^2(\Theta_0)}\,.
\end{equation} 
This expression allows us to distinguish three cases depending on the
value of $v(\Theta_0)$. For $v(\Theta_0)= 0$, the background is
obviously Minkowski, while for $v(\Theta_0)< 0$ or $v(\Theta_0)>0$, it
becomes de Sitter (dS) or Anti-de Sitter (AdS), respectively. While an
AdS scenario can be excluded on purely phenomenological grounds, the
dS case could potentially lead to a late--time acceleration of the
Universe in agreement with the observations. Note, however, that a
scale-invariant theory with spontaneous symmetry breaking contains
always a massless Goldstone mode, which is known to generate
instabilities as far as dS is
concerned~\cite{Allen:1987tz,Jalmuzna:2011qw}; see also
Refs.~\cite{Antoniadis:1985pj,Tsamis:1992sx,Tsamis:1994ca,
Antoniadis:2006wq,Polyakov:2009nq,Dvali:2013eja,Dvali:2014gua,
Dvali:2017eba,Wetterich:2017ixo, Dvali:2018fqu,Kehagias:2018uem}. We
are therefore left with a unique scenario that might be
phenomenologically viable, namely the one in which the \textit{induced
cosmological constant} following from the potential $U(\Theta_0)$ is
appropriately~\emph{fine-tuned} to be zero by requiring
\begin{equation}\label{cond1}
  v( \Theta_0)=v'(\Theta_0)=0\,.
\end{equation}
Note that, although we set $\Lambda_0=0$ in the above derivation for
the sake of simplicity, this is not a necessary condition. Indeed,
even if $\Lambda_0\neq 0$, the late time evolution of the Universe
will be eventually dominated by a constant component if the condition
\eqref{cond1} is not satisfied, giving rise to an eternal de Sitter
expansion and to the resurgence of instabilities.

Reinserting the $\Lambda_0$ contribution, the
Lagrangian~(\ref{MSattractor}) at the minimum \eqref{cond1} boils down
to
\begin{equation}\label{MSattractor4} 
\frac{\mathscr L}{\sqrt{g}}\simeq
\frac{M_P^2}{2}R -\frac{1}{2}\Theta_{0}(\partial
\Phi)^2 - \Lambda_0\, K_\Lambda(\Theta_{0}) \,e^{-4 \Phi/M_P }\ ,
\end{equation}
which must be supplemented with that of the particles produced during
the heating stage.  If $\Lambda_0 K_\Lambda(\Theta_{0})>0$, the
potential term in this expression is of a runaway type. In order not
to overclose the Universe, the energy density in the dilaton field
should be rather small, namely
\begin{equation} 
\frac{1}{2} \Theta_{0}(\partial_0\Phi)^2+\Lambda_0\,
K_\Lambda(\Theta_{0}) \,e^{-4 \Phi/M_P } \lesssim 10^{-120}M_P^4\ ,
\end{equation} 
with the right--hand side of the above inequality standing for the
present critical energy density. Given this restriction, the expansion
rate of the Universe will be initially dominated by the radiation and
matter components generated during the heating stage.  The field
$\Phi$ behaves essentially as a thawing quintessence field
\cite{Wetterich:1987fm,Ratra:1987rm,Caldwell:2005tm,Ferreira:1997hj}. In
particular, it stays frozen at the value $\Phi_0$ inherited from
inflation till the moment in which the decreasing energy density of
the heating products becomes comparable to its approximately constant
energy density. When that happens, the dilaton starts rolling towards
$\Phi\to\infty$, while driving the present--day accelerated expansion.

\subsection{A worked--out example}
\label{sec:example}

To illustrate the cosmological consequences of the general Lagrangian
density \eqref{MSattractor}, we will restrict ourselves to a simple
scenario involving a maximally--symmetric \textit{hyperbolic}
field--manifold ($\kappa<0$) and the following set of potentials
\begin{equation}\label{eq:workout}
U(\Theta)=U_0\left(1-\frac{\Theta}{\Theta_{0}}\right)^2\
,\hspace{20mm}  K_\Lambda(\Theta)=\Theta^2\ .
\end{equation}
This choice is done for illustrative purposes only. Indeed, as should
be clear from Fig.~\ref{fig:stretch}, the field stretching in the
vicinity of the kinetic poles makes the observables almost insensitive
to the details of the potentials as long as inflation is concerned
~\cite{Karananas:2016kyt}. Interestingly, the constant $\Theta_0$
denoting the position of the $\Theta$--minimum in this example can be
reabsorbed into the definition of the dilaton $\Phi$. Indeed, by
performing a transformation
\be{dil-norm}
\Phi \rightarrow \gamma \Phi \ , \hspace{20mm} \Theta \rightarrow
\gamma^{-2} \Theta \ , \hspace{20mm} c\rightarrow \gamma^{-2} c\ ,
\ee
with $\gamma\equiv \Theta_{0}^{-1/2}$, we obtain 
\begin{equation}\label{MSattractor3} 
\frac{\mathscr L}{\sqrt{g}}=\frac{M_P^2}{2}R 
-\frac{1}{2}\Bigg[-\frac{M_P^2(\partial  \Theta)^{2}}{4\, \Theta
(\kappa \Theta+c)}  +\Theta (\partial\Phi)^2 \Bigg]
-U(\Theta)-U_{\Lambda_0}(\Theta,\Phi)\ , 
\end{equation}
with
\begin{equation}
\label{eq:workout2}
U(\Theta)=U_0\left(1-\Theta\right)^2\ ,\hspace{20mm}
U_{\Lambda_0}(\Theta,\Phi)\equiv \frac{\Lambda_0}{\gamma^4}\, \Theta^2  
\,e^{-4 \gamma \Phi/M_P }\ .
\end{equation} 
Written in this form, the dilaton field $\Phi$ becomes canonically
normalized at late times (i.e. when $\Theta\to 1$).

\subsubsection{Inflation}

\begin{figure}
{\begin{center}
\includegraphics[scale=0.3]{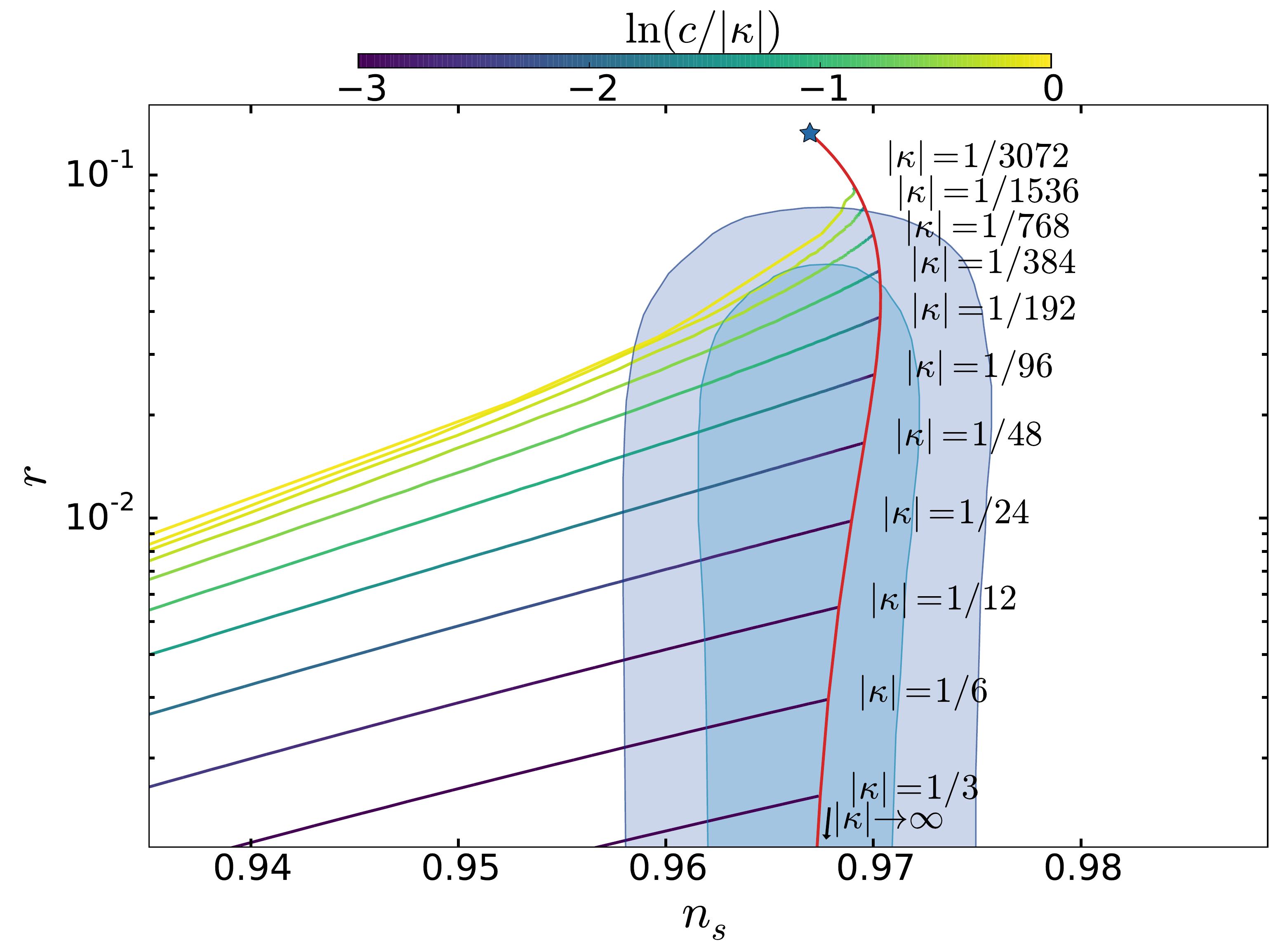}
\end{center}}
\caption{The ($n_s,r$)--plane for a biscalar model with
maximally--symmetric hyperbolic kinetic sector and the
theory--defining functions in~(\ref{eq:workout}). The red line follows
from the exact analytic expressions in Eqs.~(\ref{nsc0})
and~(\ref{rc0}), corresponding to the $c=0$ case. This line
interpolates between the chaotic $m^2\phi^2$ inflationary
predictions~(\ref{appsmallk}) at small $\vert\kappa\vert$, and the
Higgs/Starobinsky inflation predictions~(\ref{applargek}) at large
$\vert\kappa\vert$. The shaded regions mark the Planck 2018
constraints at 68\% and 95\% C.L. obtained for a $\Lambda$CDM
model~\cite{Akrami:2018odb}. As evident from the plot, the bounds on
the tensor--to--scalar ratio constrain the field-space curvature
$\kappa$.}
\label{fig:ns_r}  
\end{figure}

As argued in the previous section, for a phenomenologically viable
choice of $\Lambda_0$, both the symmetry breaking term $U_{\Lambda_0}$
and the dilaton field $\Phi$ can be safely neglected during the
inflationary stage. We are, therefore, left with a single $\Theta$
component, whose scalar and tensor perturbations can be computed using
the standard techniques. To this end, we parametrize the spectra of
these fluctuations in the almost scale-invariant
form~\cite{Mukhanov:1990me}
\begin{equation} 
\label{defspec}
P_s=A_s\left(\frac{k}{k_*}\right)^{n_s-1+\frac12\alpha_s \ln
(\frac{k}{k_*})}\ ,\hspace{20mm}
P_t=A_t\left(\frac{k}{k_*}\right)^{n_t}\ ,
\end{equation}
and compute the inflationary observables
\begin{eqnarray} 
\label{eq:infl_obser}
A_s&=&\frac{1}{24\pi^2 M_P^4}\frac{U}{\epsilon}\ , \hspace{35mm}
n_s=1+ 2\eta-6\epsilon\ ,\label{Ansdef} \\
\alpha_s&=&8\epsilon(2\eta-3\epsilon)-2\delta^2\ , \hspace{25mm}
r\equiv \frac{A_t}{A_s}=-8n_t=16\epsilon \label{asrdef}\ .
\end{eqnarray}
In the above we have introduced the standard slow-roll parameters, but
appropriately adapted to the non-canonical scalar field $\Theta$,
\begin{equation}
\label{epsdef}
\epsilon\equiv\frac{M_P^2}{2 K}\left(\frac{U_{,\Theta}}{U}\right)^2\ ,
\hspace{7mm} \eta \equiv
\frac{M_P^2}{\sqrt{K}U}\left(\frac{U_{,\Theta}}{\sqrt{K}}\right)_{,\Theta}      
\ , \hspace{7mm} \delta^2 \equiv \frac{M_P^4 U_{,{\Theta}}}{ K
U^2}\left[\frac{1}{\sqrt{K}}
\left(\frac{U_{,\Theta}}{\sqrt{K}}\right)_{,\Theta}\right]_{,\Theta}\ ,
\end{equation} 
where $K\equiv K(\Theta)$ can be read from Eq.~(\ref{req_gho_free_2})
or equivalently from the Lagrangian densities (\ref{MSattractor}) and
(\ref{MSattractor3}).  The quantities $A_s$, $n_s$, $\alpha_s$ and $r$
in~(\ref{eq:infl_obser}) and \eqref{asrdef} should be understood as
evaluated on the field value $\Theta_*\equiv \Theta(N_*)$, at which
the reference pivot scale $k_*$ in Eq.~(\ref{defspec}) exits the
horizon, or in other words at $k_*=a_* H_*$. Here,
\begin{equation}\label{Nefolds}
N_*=\frac{1}{M_P}\int_{\Theta_{\rm E}}^{\Theta_*}\frac{\sqrt{K}d
\Theta}{\sqrt{2\epsilon}}=\frac{1}{8c}\ln\left[\frac{\Theta_*}{\Theta_{\rm 
E}}\left(\frac{\kappa \Theta_{\rm E}+c}{\kappa\Theta_*+c}
\right)^{1+\frac{c}{\kappa}}\right]\ ,
\end{equation}
stands for the corresponding number of inflationary $e$--folds, and 
\begin{equation}
\Theta_{\rm E}=\frac{1-4c-2\sqrt{4c^2-2c-2\kappa}}{1+8\kappa} \ ,
\end{equation}
denotes the value of the $\Theta$--field at the end of inflation. As
usual, this is defined by the condition $\epsilon(\Theta_{\rm
E})\equiv 1$. By inverting Eq.~\eqref{Nefolds}, we can express the
inflationary observables as functions of the model parameters and
$N_*$. For general values of $c$ and $\kappa$, this inversion cannot
be analytically performed and one must rely on numerical methods. The
values of the spectral tilt $n_s$ and the tensor--to--scalar ratio
$r$, following from a numerical treatment of the
potential~\eqref{eq:cantheta}, are presented in Fig.~\ref{fig:ns_r}.

The qualitative behavior of these observables can be understood by
considering two limiting cases in parameter space:
\begin{enumerate} 
\item~\emph{Quadratic pole limit}. For $c=0$, the kinetic pole in
  Eq.\eqref{MSattractor3} becomes quadratic. In this limit,
  Eq.~(\ref{Nefolds}) yields
\begin{equation}
 N_*=\mathcal N_* -\frac{1}{8\vert
  \kappa\vert}\left(\frac{1}{\Theta_{\rm E}}+\ln \Theta_{\rm
    E}\right)\ , 
\end{equation}
with 
\begin{equation}
\mathcal N_*= \frac{1}{8\vert \kappa\vert
}\left(\frac{1}{\Theta_*}+\ln\Theta_*\right) \
,~~~\text{and}~~~\Theta_E= 
\frac{1}{1+\sqrt{8\vert\kappa\vert}}\ .
\end{equation}
Using the above, it is straightforward to see that~\cite{Rubio:2018ogq} 
\begin{equation}\label{solc0}
\Theta_*=-\frac{1}{{\cal W}_{-1}\left[-e^{-8\vert\kappa \vert \mathcal
N_*}\right]} \ , 
\end{equation}
with ${\cal W}_{-1}$, the lower branch of the Lambert $\mathcal
W$--function. Inserting Eq.~(\ref{solc0}) into~(\ref{epsdef}) and
taking into account the relations (\ref{Ansdef}) and~(\ref{asrdef}),
we obtain the following analytic expressions for the amplitude of the
primordial spectrum of scalar perturbations
\be{Asc0}
A_s =\frac{U_0}{192 \,\vert\kappa\vert\, \pi^2 M_P^4}\frac{(1+{\cal
 W}_{-1})^4}{{\cal W}_{-1}^2} \ ,
\ee 
for the spectral tilt and its running 
\be{nsc0}
n_s =1-16\vert \kappa\vert \frac{1-{\cal W}_{-1}}{\left(1+{\cal
W}_{-1}\right)^2}\ , \hspace{20mm} \alpha_s = -128\,\vert \kappa\vert 
^2\frac{{\cal W}^2_{-1}-3 {\cal W}_{-1}}{(1+{\cal W}_{-1})^4} \ ,
\ee 
and, finally, for the tensor--to--scalar ratio 
\be{rc0}
r=\frac{128\, \vert\kappa \vert }{\left(1+{\cal W}_{-1}\right)^2} \ .
\ee 
Note that the above quantities are non-trivially related as
\begin{equation}
\label{eq:bound2}
r = \frac{(1-n_s)^2}{2\vert \kappa\vert Y_1^{2}} \ ,
\hspace{25mm}
\alpha_s=-\frac{1}{2}(1-n_s)^2\,Y_2\,,
\end{equation} 
with $Y_1$ and $Y_2$ given by 
\begin{equation}
\label{eq:y1y2}
Y_1\equiv \frac12 \left(1+\sqrt{1+y}\right)\ ,\hspace{20mm} Y_2\equiv
\frac{(y+2Y_1)(y+8Y_1)}{(2Y_1)^{4}}\ ,
\end{equation}
and 
\begin{equation}
y\equiv \frac{1-n_s}{2\vert\kappa\vert}\,.
\end{equation}
The inflationary observables~(\ref{nsc0}) and~(\ref{rc0}) display an
interesting attractor behavior at large $\vert\kappa\vert N_*$, very
similar to that appearing in the $\alpha$-attractor
scenarios~\cite{Ferrara:2013rsa,Kallosh:2013yoa,Galante:2014ifa}. Indeed,
by taking into account the Lambert
function bound~\cite{DBLP:journals/corr/Chatzigeorgiou16},
\begin{equation}
\mathcal W_{-1}[-e^{-8\vert\kappa\vert \mathcal N_*}]>-8\vert\kappa\vert 
\mathcal N_*-\sqrt{2(8\vert\kappa\vert \mathcal N_*-1)}  
\ ,
\end{equation}
we obtain
\begin{equation}\label{applargek}
n_s\simeq 1-\frac{2}{\mathcal N_*}\ , \hspace{20mm } r\simeq \frac{2}{
\vert\kappa\vert \mathcal N_*^2}\ , \hspace{20mm} \alpha_s=-\vert
\kappa\vert \,r \ ,
\end{equation}
at $8\vert\kappa\vert \mathcal N_*\gg 1$. In this limit---namely for
$1-n_s \ll2\vert \kappa\vert$, or equivalently $y\ll 1$---the
functions $Y_1$ and $Y_2$ approach their minimal value $Y_1=Y_2=1$, as 
can be immediately verified from~(\ref{eq:y1y2}). Consequently, we
have
\begin{equation}
\label{eq:bound3}
r = \frac{(1-n_s)^2}{2 \vert \kappa\vert}\ ,
\hspace{25mm}\alpha_s= - \frac12 (1-n_s)^2\ .
\end{equation} 
Interestingly, the tensor--to--scalar ratio approaches zero at $\vert
\kappa\vert \to \infty$.

In the opposite limit, i.e. for $\vert \kappa\vert \to 0$ (which
should of course be taken with care when $c\to 0$), the predictions
coincide with those of the $m^2\phi^2$ chaotic inflationary
scenario,\,\footnote{Note that now the expressions contain $N_*$
rather than $\mathcal N_*$.}
\begin{equation}\label{appsmallk}
n_s\simeq 1-\frac{4}{1+2N_*}\simeq 1-\frac{2}{N_*}\ , \hspace{15mm} 
r\simeq \frac{16}{1+2N_*}\simeq \frac{8}{N_*} \ ,\hspace{15mm}
\alpha_s=-\vert\kappa\vert r \ ,
\end{equation}
and  the relations in~(\ref{eq:bound2}) become
\begin{equation}
\label{eq:bound4}
r=4 (1-n_s)\ ,
\hspace{25mm}\alpha_s= -(1-n_s)^2 \ .
\end{equation}

\begin{figure}
\centering
\includegraphics[scale=0.4]{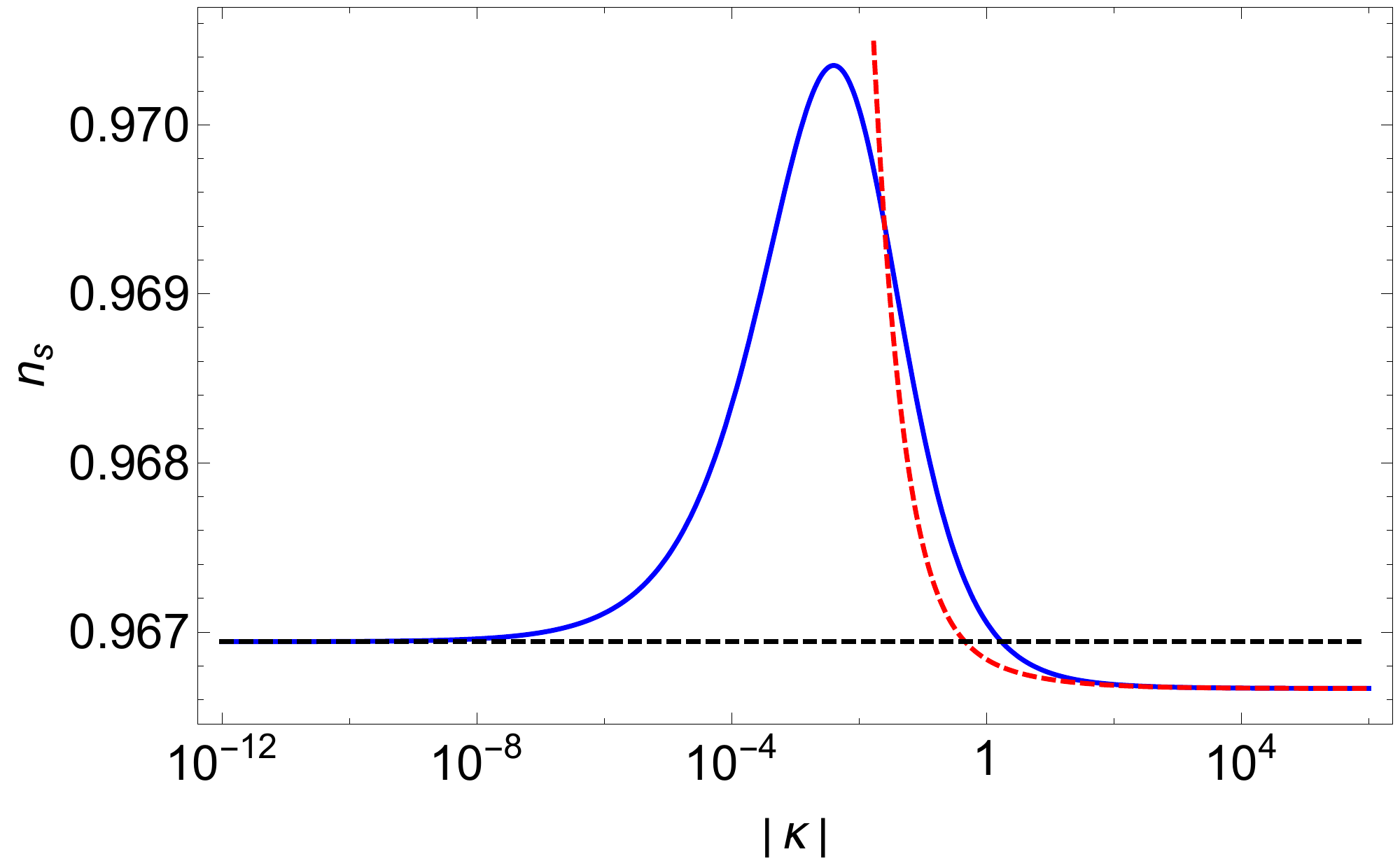}
\includegraphics[scale=0.4]{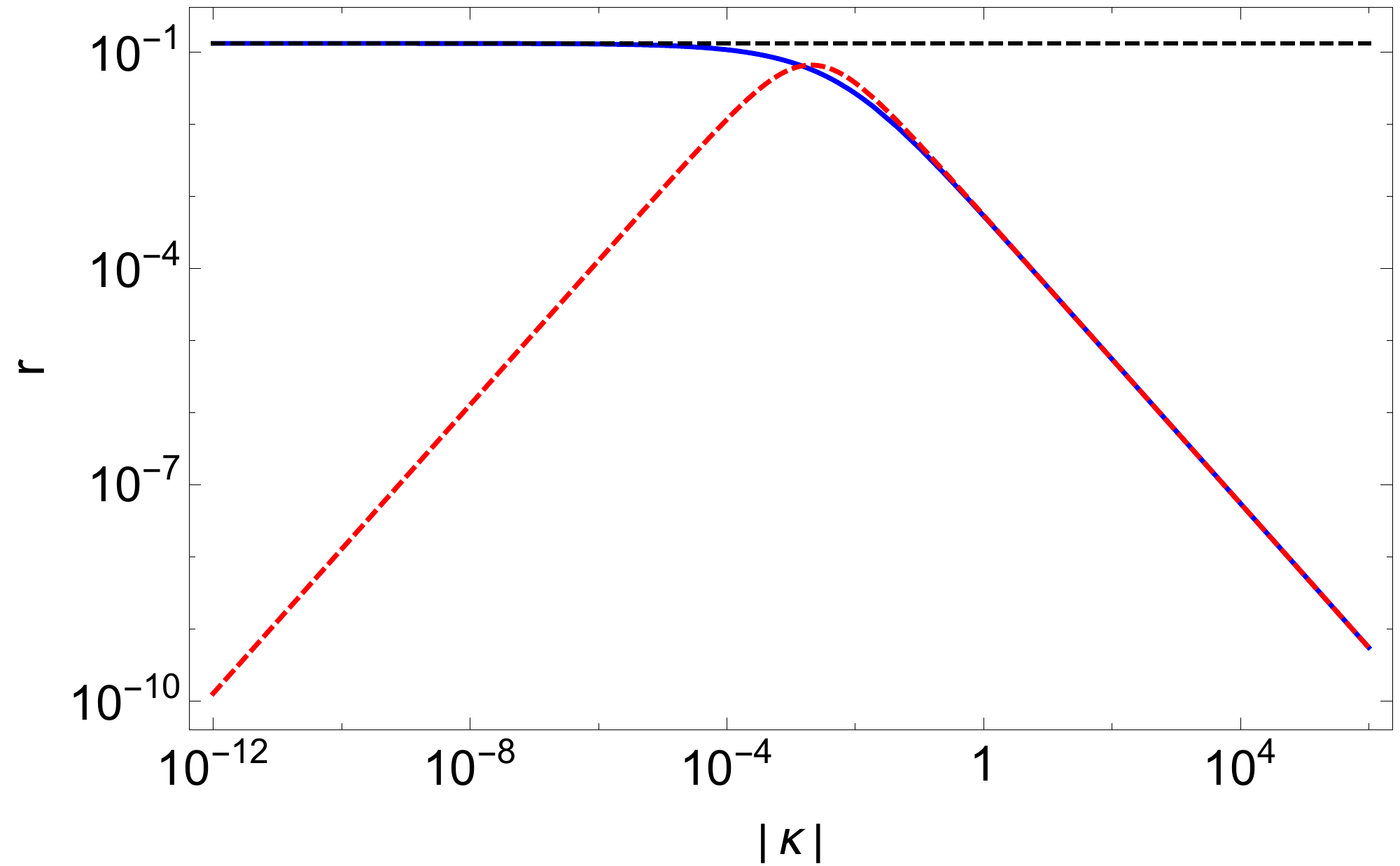}
\caption{Comparison between the approximate
expressions~(\ref{applargek}) (red dashed line) and~(\ref{appsmallk})
(black dashed line) for the spectral tilt $n_s$ and tensor--to--scalar
ratio $r$ and the most general expressions in Eqs.~(\ref{nsc0})
and~(\ref{rc0}) (blue solid line).}
\label{fig:comparison}
\end{figure}

The comparison of the approximate expressions~(\ref{applargek})
and~(\ref{appsmallk}) for the spectral tilt and tensor--to--scalar
ratio with the most general ones given in Eqs.~(\ref{nsc0})
and~(\ref{rc0}) is shown in Fig.~\ref{fig:comparison}.

\item~\emph{The quadratic--to--linear pole transition.}  For $c\neq
0$, the inflationary pole at $\Theta=0$ is no longer reachable and we
are left with a linear pole at $\Theta=c/\vert\kappa\vert$. To
understand the consequences of this pole, let us consider the
inversion of Eq.~(\ref{Nefolds}) in the limit $c/\vert \kappa\vert \ll
1$ and $4\vert \kappa\vert N_*\gg1$. We obtain
\begin{equation}
\mathcal N_*\simeq \frac{1}{8c}\ln
\frac{\Theta_*}{\Theta_*-c/\vert\kappa\vert}\hspace{10mm}\rightarrow  
\hspace{10mm}\Theta_*(\mathcal N_*)\simeq \frac{c}{\vert
\kappa\vert}\frac{e^{8 c \mathcal N_*}}{e^{8 c \mathcal N_*}-1} \ ,
\end{equation}
with 
\begin{equation}
\mathcal N_*\equiv N_* -\frac{1}{8c}\ln
\left(1-\frac{c}{\vert\kappa\vert \,\Theta_{\rm E}}\right)\ .
\end{equation}
To the lowest order in $c/\vert \kappa\vert$, the inflationary
observables become
\begin{eqnarray}\label{As}
A_s &=&\frac{\lambda \sinh^{2} \left(4 c \mathcal N_*
 \right)}{1152\pi^2 \xi^2_{\rm eff}\, c^2} \ ,\hspace{30mm} n_s =
 1-8\,c \coth\left(4 c \mathcal N_*\right)\ , \\ \alpha_s &=& - 32\,
 c^2 \sinh^{-2} \left(4 c \mathcal N_* \right)\ , \hspace{20mm} r=
 \frac{32\, c^2}{\vert\kappa\vert} \sinh^{-2} \left(4 c \mathcal
 N_*\right)\,,\label{r}
\end{eqnarray}
where we introduced an effective coupling
\begin{equation}
\hspace{10mm} 
\xi_{\rm eff}\equiv\frac{1}{\sqrt{6 a^2 \vert\kappa\vert}}\ ,
\end{equation}
and defined
\be{}
a=-\frac{1-6|\kappa|}{|\kappa|} \ .
\ee

For $4 c \mathcal N_*>1$, the spectral tilt decays linearly and the
tensor--to--scalar ratio approaches zero asymptotically, i.e.
\begin{equation}
n_s\simeq 1-8 c\ , \hspace{20mm } r\simeq 0\ . 
\end{equation}
\end{enumerate}

\subsubsection{Dark energy}

After the end of inflation, the field $\Theta$ will perform
oscillations around the minimum of its effective potential, heating
the Universe and eventually relaxing to $\Theta=1$. When that happens,
the Lagrangian boils down to
\begin{equation}\label{MSattractor5} 
\frac{\mathscr L}{\sqrt{g}}\simeq
\frac{M_P^2}{2}R -\frac{1}{2}(\partial
\Phi)^2 - \frac{\Lambda_0}{\gamma^4}\, \,e^{-4\gamma \Phi/M_P }\ . 
\end{equation}
It is therefore clear that the dilaton can drive an accelerated
expansion of the Universe for suitable values of $\Lambda_0$ and
$\gamma$. At early times, the potential of $\Phi$ is small as compared
to the Hubble rate. This prevents the motion of the field and forces
it to stay frozen at the value that it inherited from inflation. At
late times, the dilaton starts evolving and the system approaches an
effective equation--of--state parameter
\cite{Wetterich:1987fm,Copeland:1997et,Scherrer:2007pu}
\begin{equation}\label{eqn:eos_omega_quintessence}
1+w= \frac{16 \gamma^2}{3} F(\Omega_{\rm DE})\
,\hspace{7mm}\text{with}\hspace{7mm}F(\Omega_{\rm DE})=\left[\frac{1}{
   \sqrt{\Omega_{\rm
DE}}}-\Delta\tanh^{-1} \sqrt{\Omega_{\rm DE}}\right]^2\ ,
\end{equation}
which leads to acceleration ($w<-1/3$) if $\gamma<1/(2
\sqrt{2})$. Here, 
\begin{equation}
\Omega_{\rm DE}\equiv\frac{1}{1+\Delta_{0}\,a^{-3}}\ ,
\end{equation}
stands for the dark energy abundance associated with the dilaton
component and  
\begin{equation}
\Delta  \equiv \frac{1-\Omega_{\rm DE}}{\Omega_{\rm DE}}\
,\hspace{20mm}  
\Delta_0  \equiv \frac{1-\Omega_{\rm DE,0}}{\Omega_{\rm DE,0}}\ ,
\end{equation}
where the subscript $0$ refers to quantities evaluated today.

\subsubsection{Connecting inflation and dark energy}
\label{sec:connection_infl_de}

Until this point, we have assumed that the parameters $\kappa$, $c$
and $\gamma$ in our example are independent. If these quantities were
related, the set of scale--invariant maximally--symmetric TDiff
theories will also lead to non--trivial connections between the
inflationary and DE observables. This is what happens for instance in
the simplest scale--invariant model that can be constructed out of two
scalar fields $\phi_1$ and $\phi_2$, namely~\cite{Shaposhnikov:2008xb}
\begin{equation}
\label{eqn:hdm_lagrangianHD}
\frac{\cal L}{\sqrt{g}}=\frac{\xi_1 \phi_1^2 +\xi_2 \phi_2^2}{2} R
-\frac{1}{2} (\partial \phi_1)^2 -\frac{1}{2} (\partial
\phi_2)^2-\frac{\lambda}{4} \left(\phi_1^2-\alpha\phi_2^2 \right)^2
+\Lambda_0\ ,
\end{equation}
with $\xi_1,\xi_2,\lambda$ and $\alpha$ positive dimensionless
couplings and $\Lambda_0$ constant. This Diff-equivalent Lagrangian 
density follows from the TDiff-invariant one in (\ref{Tdiffinit})
after restoring the full symmetry with the St\"uckelberg trick of
Sec.~\ref{sec:scal_inv_two_field} and
with the following choice of theory--defining functions (see
also Ref.~\cite{Blas:2011ac} for more examples)
\begin{equation}
\begin{aligned}
&G_1(g)= \beta^2\,g^{2(\beta-1)} \ ,\hspace{10mm} ~~G_2(g)=
\beta\,g^{2\beta-1} \ ,\hspace{10mm} G_3(g)= 1+g^{2\beta}\ ,
\\ &f(g)=\xi_1+\xi_2\,g^\beta\ ,\hspace{12mm} ~~v(g)=
\frac{\lambda}{4}\left(1-\alpha g^{2\beta}\right)^2 \ .
\end{aligned}
\end{equation}
Here, $\beta$ is an arbitrary constant, and to
obtain~(\ref{eqn:hdm_lagrangianHD}), we have identified $\phi=\phi_1$
and introduced $\phi_2=\phi_1\,g^\beta$. When transformed to the
Einstein frame and rewritten in terms of variables
\begin{equation}
\gamma^{-2} \Theta \equiv
\frac{(1+6\xi_1)\phi_1^2+(1+6\xi_2)\phi_2^2}{\xi_1
\phi_1^2+\xi_2\phi_2^2}\ , \hspace{10mm}
\exp\left[\frac{2\gamma\Phi}{M_P}\right] \equiv
\frac{\kappa_c}{\kappa}
\frac{(1+6\xi_1)\phi_1^2+(1+6\xi_2)\phi_2^2}{M_P^2}\ ,
\end{equation}
with
\begin{equation}
\gamma \equiv \sqrt{\frac{\xi_2}{1+6\xi_2}}\  ,
\end{equation}
the Lagrangian density~(\ref{eqn:hdm_lagrangianL})
approximately\footnote{The main difference is associated to an
additional pole $\Theta=1$ in the Einstein-frame kinetic sector of
\eqref{eqn:hdm_lagrangianHD}. This ``Minkowski'' pole is, however,
irrelevant for the cosmological phenomenology discussed in this paper,
for details
cf. Refs.~\cite{Karananas:2016kyt,Casas:2017wjh,Rubio:2018ogq}.}
reduces to the form~(\ref{MSattractor3}) \cite{Casas:2017wjh}, with
$U(\Theta)$ and $K_\Lambda(\Theta)$ given by Eq.~(\ref{eq:workout})
and
\begin{equation}\label{kappadef}
U_0\equiv\frac{\lambda M_P^4}{4}\left(\frac{1+6
\kappa}{\kappa}\right)^2 \ ,\hspace{8mm} \kappa
\equiv\kappa_c\left(1-\frac{\xi_2}{\xi_1}\right)\ ,\hspace{8mm}
\kappa_c \equiv-\frac{\xi_1}{1+6\xi_1}\ , \hspace{8mm}
c\equiv\frac{\kappa}{\kappa_c}\gamma^2\ .
\end{equation}
A simple inspection of these expressions reveals that the parameters
$\kappa$, $c$ and $\gamma$ in this particular scenario are not
independent.  This allows us to obtain a set of consistency relations
among the inflationary and dark energy observables. An analytic form
for these consistency relations can be obtained in the limit
$\vert\kappa\vert\approx \vert\kappa_c\vert$, corresponding to an
inflationary dynamics essentially dominated by the $\phi_1$ component,
i.e. with $\xi_1\gg \xi_2$. Indeed, combining the
expression~(\ref{eqn:eos_omega_quintessence}) with those for the
spectral-tilt, its running, and the tensor-to-scalar ratio in
Eqs.~(\ref{As}) and~(\ref{r}), we obtain~\cite{Casas:2017wjh}
\be{nswcons}
n_s = 1- \frac{2}{{\cal N}_*}X \coth X\ ,\hspace{15mm}
r=\frac{2}{\vert\kappa_c \vert {\cal N}_*^2} X^2\sinh^{-2} X\
,\hspace{15mm} \alpha_s=-\vert \kappa_c\vert r\ ,
\ee
with
\be{}
X\equiv 4c{\cal N}_*=\frac{3 {\cal N}_*(1+w)}{4
F(\Omega_\textrm{DE})}\ .
\ee
Given the value of $\lambda$ at the inflationary scale, the constant
$\vert \kappa\vert$ can be determined from the amplitude of the
primordial power spectrum~(\ref{As}). For not too small values of
$\lambda$, the effective coupling is typically rather large, $\xi_{\rm
eff}\simeq \xi_1\gg 1$, leading to values of $\vert\kappa\vert$ very
close to $1/6$. In this limit, the expressions in~(\ref{nswcons})
reduce to those first found in the context of the~\emph{Higgs-dilaton
model}~\cite{GarciaBellido:2011de}.

\section{Comparison with present data sets}
\label{sec:confront}

To interpret the existing data in the light of scale-invariant models,
we perform a Markov Chain Monte Carlo (MCMC) analysis similar to those
in Refs.~\cite{Trashorras:2016azl,Casas:2017wjh}. In particular, we
sample the posterior probability distribution $P=p(\theta | x, M)$ of
cosmological parameters $\theta$ given the data $x$ and a model $M$ by
means of the Bayes theorem
\be{} 
P = \frac{ p(\theta |
M)}{E} \, \mathcal{L} \ , 
\ee 
with $p(\theta|M)$ the prior distribution of parameters, given the
model and $\mathcal{L}=p(x|\theta, M)$ the likelihood. The evidence
$E=p(x|M)=\int{\mathrm{d}\theta {p(x | \theta, M) p(\theta | M) }}$
follows as a normalization factor. Once the likelihood and the priors
are given, the MCMC algorithm constructs a chain of points whose
density is proportional to the posterior probability distribution
$p(\theta| x,M)$. For the likelihood, we include the following
observational data sets:\footnote{ We assume these data sets to be
independent and do not model cross-correlations among them.}
\begin{enumerate}[i)]
\item the 2015 Planck high--multipole $TT$
likelihood~\cite{Aghanim:2015xee},\footnote{The 2018 Planck likelihood
is not yet publicly available.}
\item the 2015 Planck low--multipole polarization and temperature
likelihoods~\cite{Aghanim:2015xee},
\item the 2015 Keck/Bicep2 likelihood data release~\cite{bicep2},
\item the Joint Lightcurve Analysis data~\cite{betoule:jla},
\item the baryon acoustic oscillation data from 6dF, BOSS LOWZ, BOSS
  CMASS and SDSS~\cite{Beutler:2011hx, anderson:boss, Ross:2014qpa}.
\end{enumerate}

For the maximally--symmetric scale--invariant models under
consideration we vary $c$ and $-\ln(|\kappa|)$ in the range $[0,1]$
and $[-\ln(1/6), 8]$, with the logarithmic parametrization chosen only
for numerical convenience and the intervals motivated by the
particular example in Section \ref{sec:connection_infl_de}. For each
pair of values, we numerically solve the inflationary trajectory and
compute the spectral tilt $n_s$ and the tensor--to--scalar ratio $r$.

While the details of the heating stage after inflation remain to be
specified, here we adopt a conventional estimate that turns out to be
reasonable in many heating
scenarios~\cite{Bezrukov:2008ut,GarciaBellido:2008ab,
Repond:2016sol}. In particular, we restrict the number of inflationary
$e$--folds to a Gaussian distribution with mean $60$ and standard
deviation $2.5$. Additionally, we vary the customary cosmological
parameters using flat and non-restricting priors. The prior ranges can
be found in Table 1 of Ref.~\cite{Casas:2017wjh}.

\subsubsection*{Maximally--symmetric model without consistency
  conditions}

\begin{figure}
{\begin{center}
\includegraphics[scale=0.29]{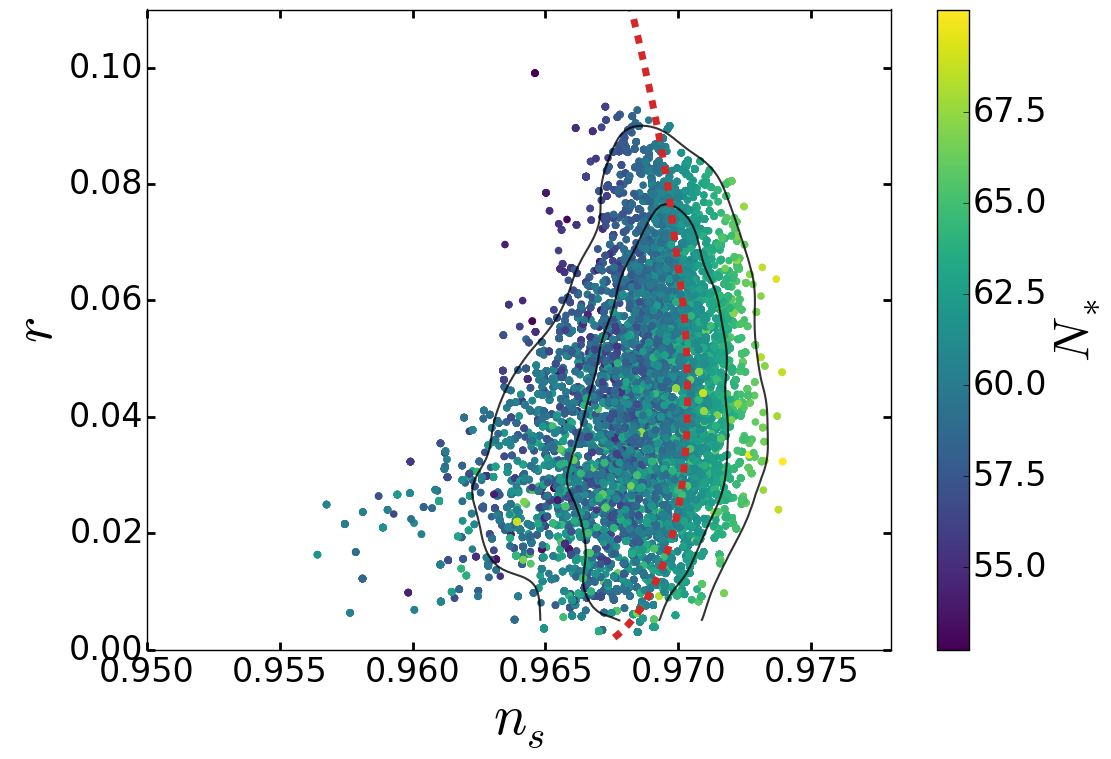}
\includegraphics[scale=0.29]{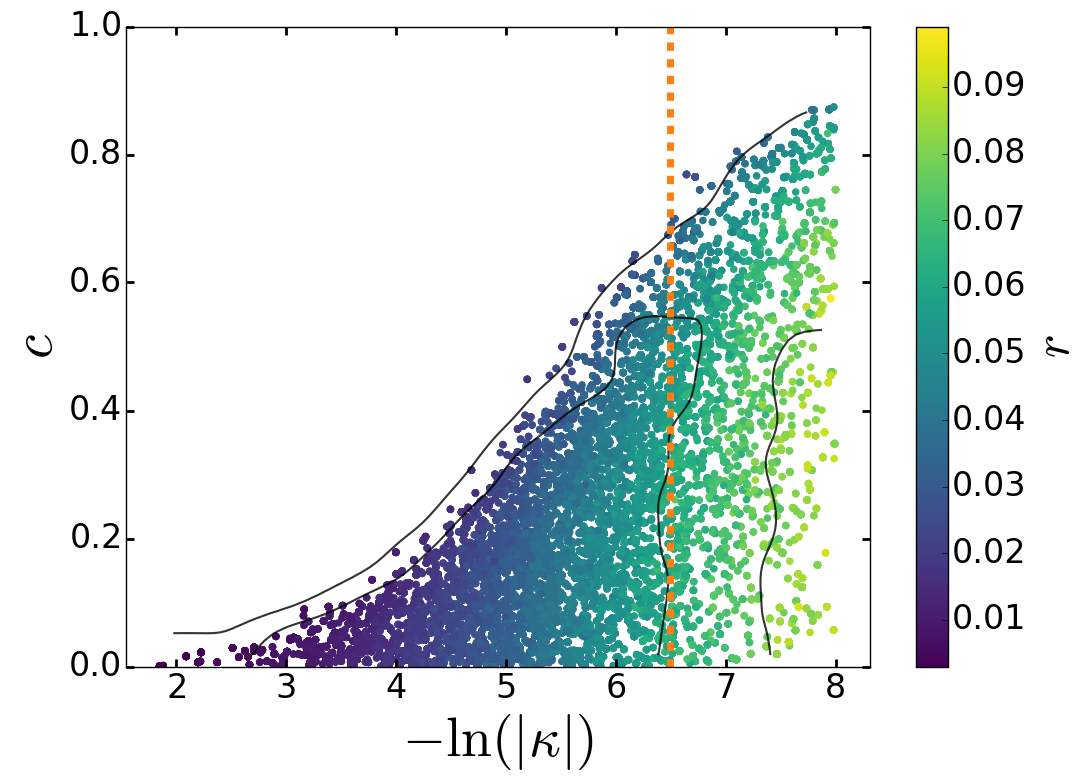}
\end{center}}
\caption{ Left: MCMC samples for the spectral tilt $n_s$ and
tensor--to--scalar ratio $r$ in a scale--invariant model~\emph{without
consistency relations}. The color coding indicates the number of
$e$-folds $N_*$ to the end of inflation. The red dashed line
corresponds to the limit $c \to 0$ for $N_*=60$.  Right: MCMC samples
for the model parameters $-\ln(|\kappa|)$ and $c$ in the same
scenario. The color coding indicates now the tensor--to--scalar ratio
$r$. For a fixed value of $\kappa$, the parameter $c$ is tightly
constrained by the spectral tilt. Note that small values of
$-\ln(|\kappa|)$ are permitted only for tiny values of $c$. This tail
corresponds to the bottom--left corner in the $r-n_s$ plot, which is
not properly explored in our parametrization. Black lines mark the
$68\%$ and $95\%$ C.L. regions. The orange dashed line corresponds to
the expected $95\%$ bound following from the Planck/BICEP 2018 data
release.  }
\label{fig:ns_r_mcmc}  
\end{figure}

To discuss how the model parameters can be constrained by \textit{pure
inflationary physics} we first study a particular realization of
\eqref{MSattractor3} with $c$ and $\kappa$ completely unrelated to
$\gamma^2$. In other words, we assume the inflationary and
dark--energy dominated eras to be completely independent. The results
of the MCMC analysis for this particular scenario are presented in
Fig.~\ref{fig:ns_r_mcmc}, both in terms of the parameters $c$ and
$\kappa$ and in terms of the observable quantities $n_s$ and $r$. As
evident from this figure, the allowed values for the spectral tilt and
the tensor-to-scalar ratio mostly correspond to a restricted version
of $\Lambda$CDM, with the curvature of the Einstein-frame kinetic
sector closely related to $r$ and the parameter $c$ constrained by the
spectral tilt $n_s$ for fixed $\kappa$.  The mean values of these
parameters are
\be{}
n_s = 0.9686_{-0.0015}^{+0.0026}\ ,\hspace{8mm} r=
 0.040_{-0.025}^{+0.016}\ ,\hspace{8mm}  c = 0.23_{-0.23}^{+0.06}  
\ ,\hspace{8mm}  -\ln(|\kappa|)=
5.5_{-1.1}^{+1.4}\ ,
\ee
with the errors denoting the $68\%$ C.L. We emphasize that these
constraints should be taken with a grain of salt for two reasons. On
the one hand, our parametrization in terms of $c$ and $-\ln(|\kappa|)$
is not suitable for large $|\kappa|$ values. On the other hand, given
the present data sets, it turns out to be quite challenging to
numerically explore the $|\kappa|\to 0$ limit at relatively large $c$
values, or, correspondingly, the region of large tensor--to--scalar
ratios and small spectral tilts. As it can be seen from the contours
in Fig.~\ref{fig:ns_r_mcmc}, the viability of relatively large
tensor--to--scalar ratios still prevents us from identifying the full
95\% confidence region for $c$ and $\kappa$. However, this is expected
to improve significantly with the eventual release of the Planck 2018
likelihood.

Both the Planck 2018 likelihood and other future CMB experiments are
expected to set tight bounds on the Einstein-frame kinetic
curvature. In particular, a decreasing limit on the tensor--to--scalar
ratio would directly translate into an increasing lower bound on
$|\kappa|$.  This becomes apparent when one considers, for instance,
the latest bound on $r$, namely $r<0.064$
\cite{Akrami:2018odb}.\footnote{Note that this bound was derived in a
$\Lambda$CDM cosmology and it should be re-evaluated for
scale-invariant models after the eventual release of the Planck 2018
likelihood.}  This value translates into a restriction $-\ln(|\kappa|)
< 6.5$, excluding therefore a large part of the Planck 2015 $(c,
\kappa)$ parameter space. Note, however, that no upper bound of
$\vert\kappa\vert$ follows from present data sets. Indeed, only an
eventual detection of primordial gravitational waves could provide an
upper limit on it.

\subsubsection*{Maximally--symmetric model with consistency
  conditions}

\begin{figure}
{\begin{center}	
\includegraphics[scale=0.28]{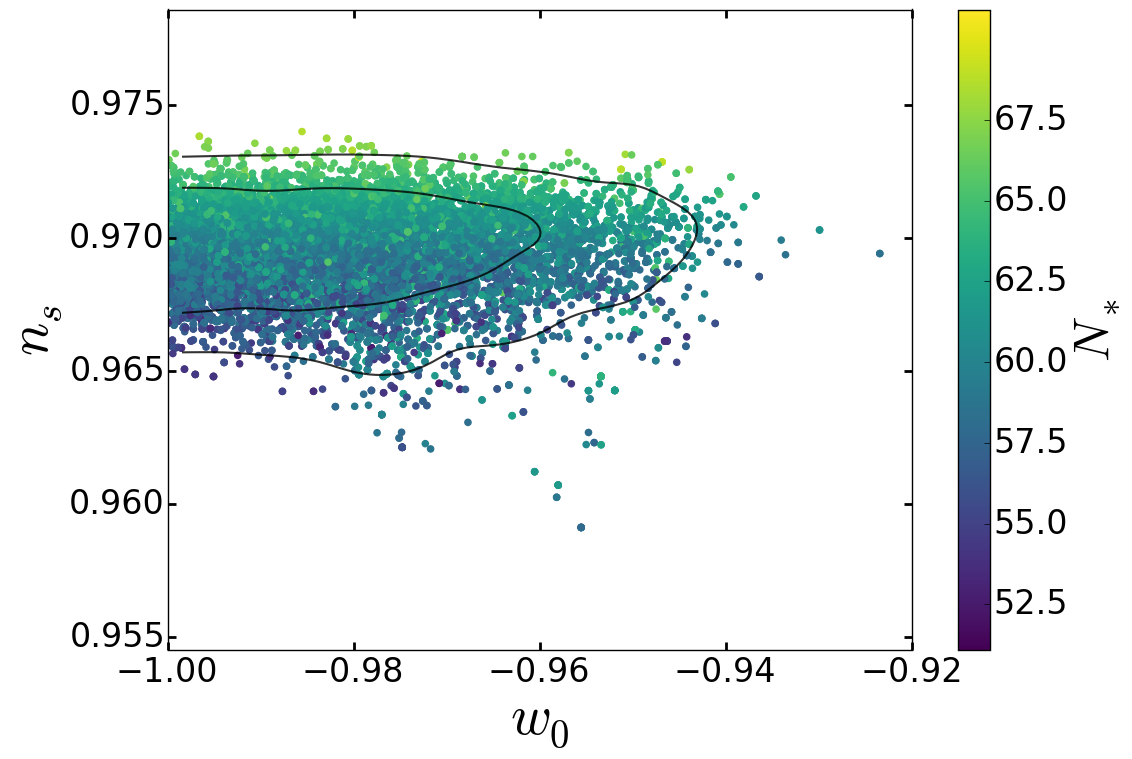}
\includegraphics[scale=0.28]{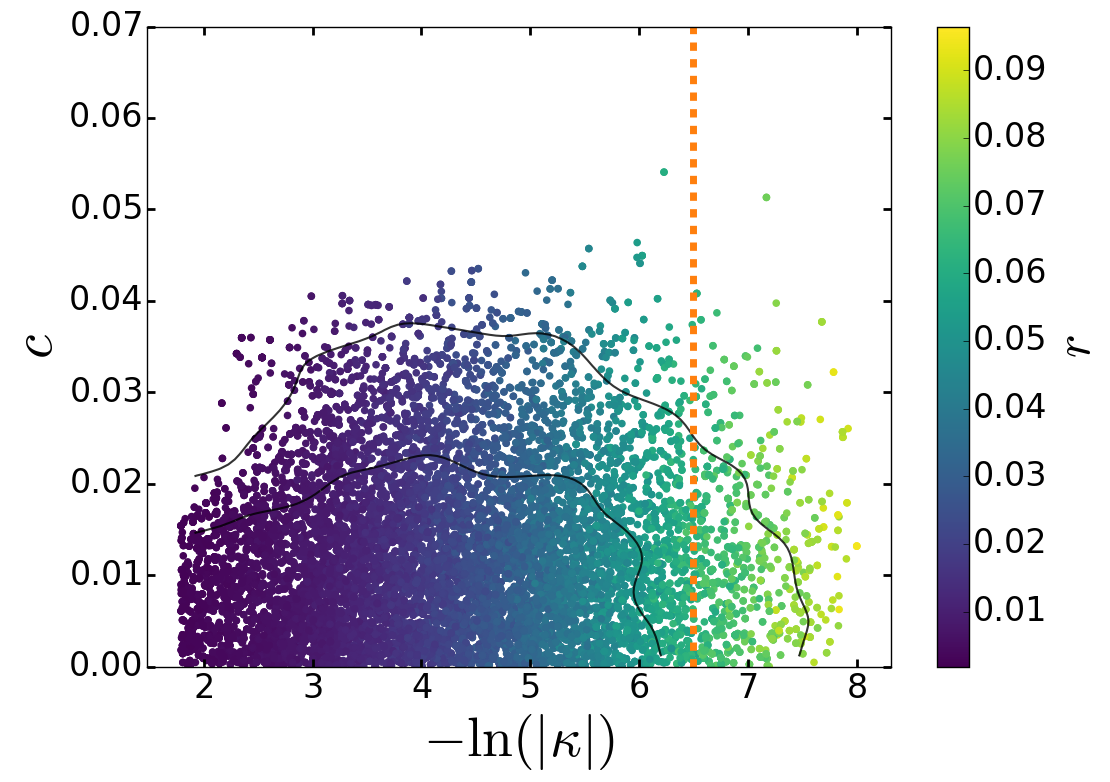}
\end{center}}
\caption{ Left: MCMC samples for the spectral tilt $n_s$ and the
equation--of--state parameter $w_0$ in a scale--invariant model
\textit{with the consistency relation $\gamma^2(c)=c$}. The color
coding indicates the number of $e$--folds $N_*$ to the end of
inflation.  Measurements of the equation--of--state parameter
constrain strongly the spectral tilt.  Right: MCMC samples for the
model parameters $-\ln(|\kappa|)$ and $c$ in the same scenario. The
color coding now indicates the tensor--to--scalar ratio $r$, while the
orange dashed line marks the expected 95\% C.L bound following from
the Planck/BICEP 2018 data release.  Note that if the early and late
Universe observables are related by a consistency relation, the
constraints on $c$ are much tighter than in the absence of it (note
the different scale for $c$ in Fig.~\ref{fig:ns_r_mcmc}).}
\label{fig:ns_w0_mcmc}  
\end{figure}

To illustrate the impact of a potential connection between the early
and the late Universe we consider now a realization
of~(\ref{MSattractor3}) involving a consistency relation $\gamma^2(c)=
c$. This choice is motivated by the simple biscalar scenario presented
in Section~\ref{sec:connection_infl_de} and should be understood just
as a particular example of the different consistency relations that
could appear in this type of models. As shown in
Fig.~\ref{fig:ns_w0_mcmc}, the existing constraints on the present
equation--of--state parameter effectively restrain the spectral tilt
and significantly reduce the 68\% C.L. range of for $c$ and
$\kappa$,\footnote{The cut on the left-hand side of the $(c,\kappa)$
plot is due to our prior restriction $|\kappa| < 1/6$.}
\begin{equation}
n_s = 0.9695_{-0.0013}^{+0.0019}\ ,\hspace{10mm} r=
 0.026_{-0.024}^{+0.007}\ ,\hspace{10mm} c=0.013_{-0.013}^{+0.003} \ , 
 \hspace{10mm}
-\ln(|\kappa|)=4.28_{-1.56}^{+1.27}\,.
\end{equation}
This parameter-space reduction is expected to become stronger in the
near future.  On the one hand, galaxy redshift surveys such as Euclid
or LSST will provide percent-level measurements of the dark-energy
equation--of--state parameter. On the other hand, Stage IV CMB
observers such as LiteBird will determine the value of the
tensor--to--scalar ratio with an unprecedented $10^{-3}$--$10^{-4}$
accuracy.

\subsubsection*{Bayesian evidence and correlation matrices}

To quantify how the scale-invariant models above compare to
$\Lambda$CDM we calculate the Bayes factor, defined as the evidence
ratio for a model $M$ and a $\Lambda$CDM scenario given the data $x$,
namely\,\footnote{This implicitly assumes that all models are equally
probable a priori.}
\be{}
B(M) = \frac{p(M|x)}{p(M_{\Lambda \text{CDM}}|x)}\ .
\ee
We compute this quantity from the obtained MCMC chains using the
method proposed in Refs.~\cite{Heavens:2017afc, Heavens:2017hkr} and
interpret the result according to the Kass and Raftery
scale~\cite{Kass:1995loi}, where a value $\vert \Delta \ln B \vert>3$
is understood as a strong statistical preference. As shown in
Table~\ref{tab:bayes_factor}, the scale--invariant model without
consistency relations appears to be slightly disfavoured with respect
to $\Lambda$CDM. On the contrary, scale--invariant model with
consistency relations seems to be preferred over the concordance
model. Although these quantitative results should not be taken at face
value,\footnote{In particular, the parameter basis is varied when
comparing the three models. Additionally, the prior on
$-\ln(|\kappa|)$ restricts the available parameter volume. This
renders the value of the Bayes factor prior
dependent~\cite{Heavens:2017afc,Heavens:2017hkr}.  A change of the
prior volume by a factor $\lambda$ would induce a change
$\ln\,\lambda$ in the Bayes ratio.} they illustrate two important
points. First, the strong preference for a scale--invariant model with
consistency relations over the one without them stresses the
importance of these conditions when dealing with existing and future
data sets. Second, the \textit{positive evidence} \cite{Kass:1995loi}
for the model with consistency relations over $\Lambda$CDM indicates
that scale--invariant scenarios can be on equal footing with---and in
some cases superior to---the concordance model.\footnote{ Note that a
Bayes factor $\ln\, B = 2.44$ corresponds to a relative probability of
approximately $11:1$ for the scale-invariant model over $\Lambda$CDM.}

\begin{table}
\centering
\begin{tabular}{cccc}
\hline 
  & \textbf{$\Lambda$CDM} & \textbf{Without consistency rel.}
  & \textbf{With consistency rel.}\tabularnewline
\hline 
Nr. of parameters  & $8$  & $10$  & $9$\tabularnewline
$\ln B$  & $0 $  & $-1.73$  & $2.44$\tabularnewline
\hline 
\end{tabular}
\protect\caption{\label{tab:bayes_factor}The maximum likelihood
estimate of the logarithm of the Bayes factor $\ln B(M)$ with respect
to a baseline $\Lambda$CDM model. Although the comparison remains
inconclusive, the scale--invariant model without consistency relations
appears to be slightly disfavoured with respect to $\Lambda$CDM. On
the contrary, a scale--invariant model with consistency condition
$\gamma^2(c)=c$ is preferred over the concordance model. }
\end{table}

\begin{figure}
\begin{center}
\includegraphics[scale=0.45]{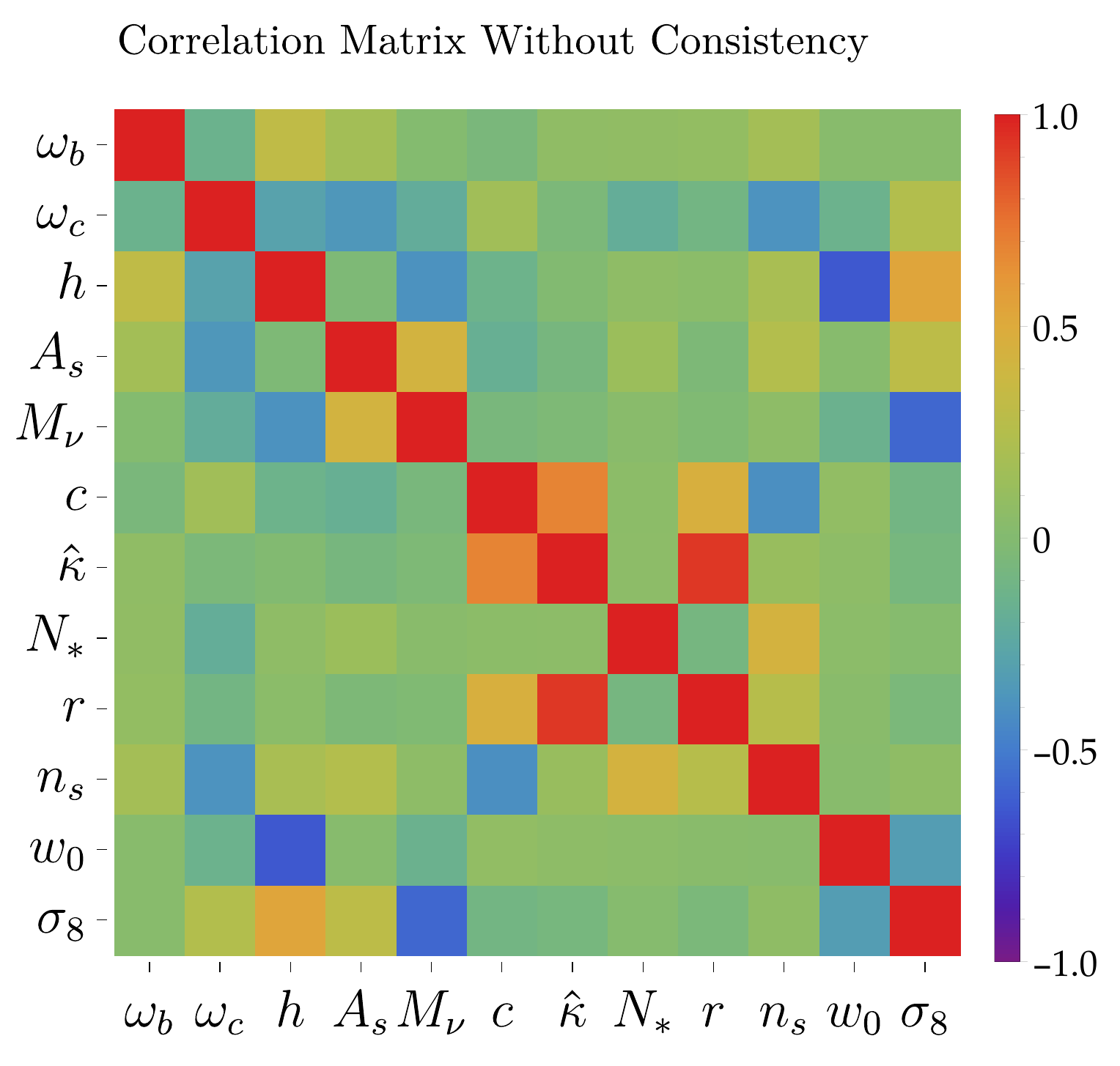}
\includegraphics[scale=0.45]{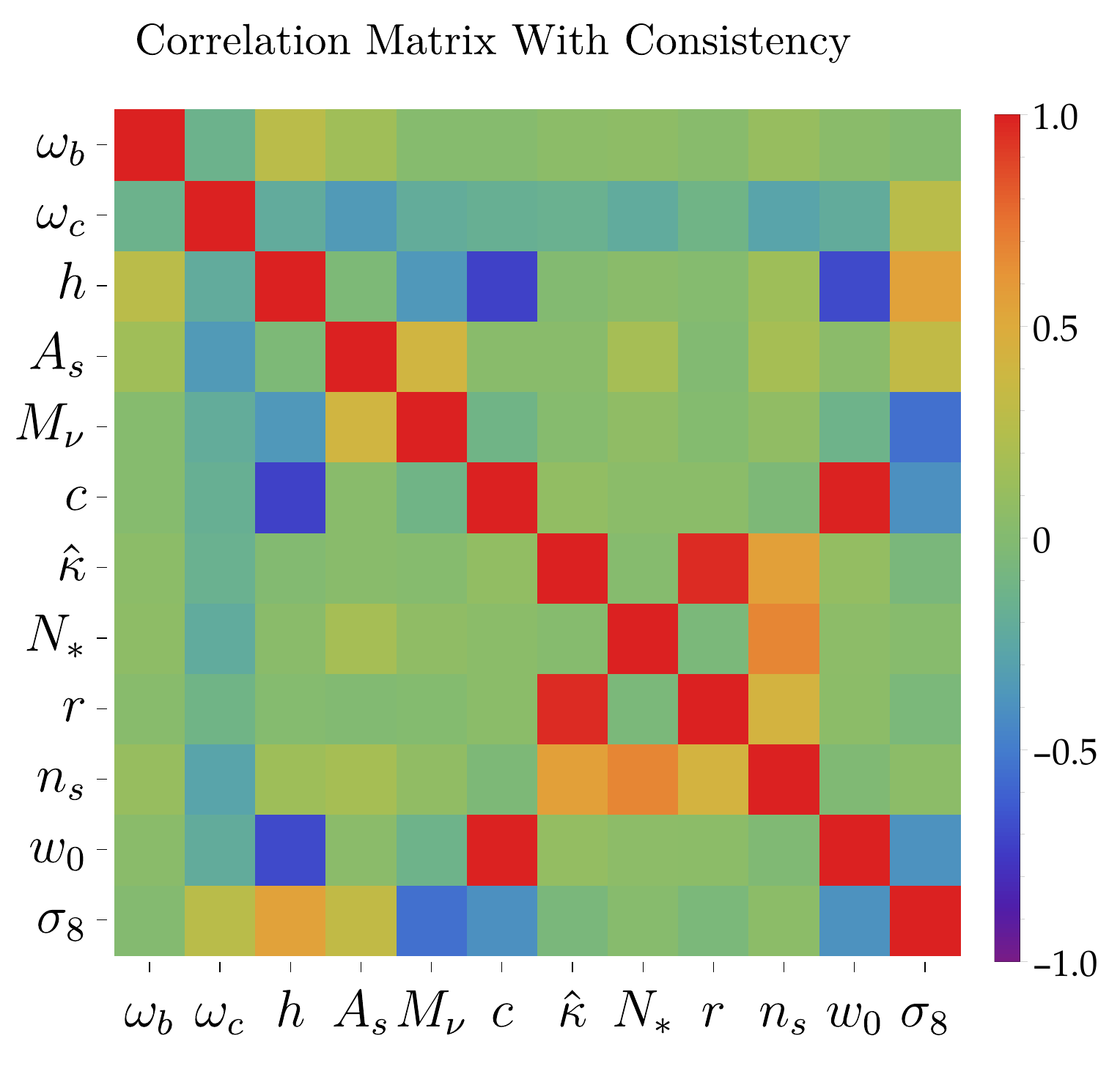} 
\end{center}
\begin{center}
\includegraphics[scale=0.45]{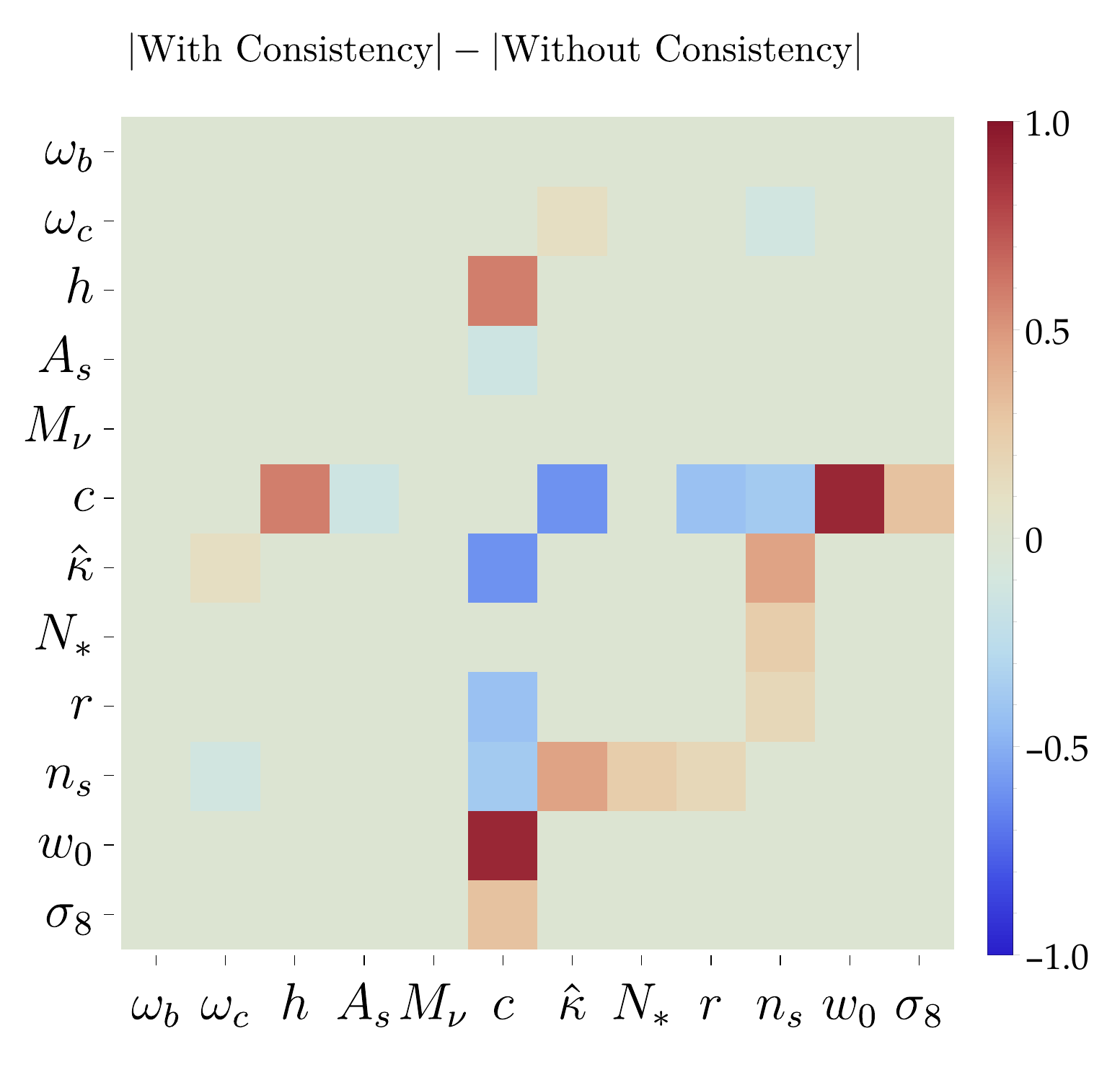}
\end{center}

\caption{(Top) Correlation matrices obtained from the covariance
matrices of the MCMC runs for a model without (left) and with (right)
consistency relations.  The $+1$ and $-1$ limits stand for totally
correlated and totally anti-correlated, respectively.  (Bottom)
Half-difference of the absolute values of the correlation coefficients
with the $+1$ and $-1$ limits indicating now if the ``correlation
strength'' has increased or decreased, respectively. In this figure we
define $\hat\kappa\equiv-\ln(\vert\kappa\vert)$ for visualization
purposes.}
\label{fig:corrMats1}  
\end{figure}

The impact of the consistency relations is also reflected in the
correlation among different cosmological parameters. This interesting
feature is shown in Fig.~\ref{fig:corrMats1}, where we display the
MCMC covariance matrices obtained from current data sets, converted
into correlation matrices. The left and right panels correspond to a
model without and with consistency relations, respectively. In these
figures we have defined $\hat\kappa\equiv-\ln(|\kappa|)$ for
visualization purposes.

Without consistency relations, there exists a positive correlation
among $c$ and $\hat\kappa$, which matches very well with the behaviour
displayed in Fig.~\ref{fig:ns_r_mcmc}, where, for a constant value of
the tensor--to--scalar ratio, an increase in $c$ corresponds to an
increase in $\hat\kappa$. In this case, the equation--of--state
parameter $w_0$ is an independent parameter, which---leaving aside the
fact that $\sigma_8$ is a derived quantity depending on all parameters
affecting the growth of structures---is only anti-correlated with the
reduced Hubble rate $h$ due to the expansion of the Universe.

When including the consistency relations, $w_0$ is no longer an
independent parameter, but rather a derived one, totally correlated
with $c$. This means that $c$ has now taken the role of a dark energy
equation--of--state parameter and, consequently, is now
anti-correlated with the reduced Hubble rate $h$. Additionally, we
observe strong positive correlations between $\hat\kappa$ and the
standard inflationary parameters $n_s$ and $r$.  Moreover, c and
$\hat\kappa$ are basically uncorrelated and independent of each
other. Both of these features are reflected in the right panel of
Fig.~\ref{fig:ns_w0_mcmc}, the former by noting that for a constant
value of $c$ the tensor--to--scalar ratio increases for increasing
$\hat\kappa$ and the latter by the observation that the $\hat\kappa$ -
$c$ contours are almost circular.

The above findings are summarized at the bottom panel of
Fig.~\ref{fig:corrMats1}, where we display the difference between the
absolute value of the correlation coefficients in the model with and
without consistency relations. The red (blue) color corresponds to
parameters which are more (less) correlated with (without) consistency
relations. In the presence of consistency relations, we see three main
features: i) $c$ and $\hat\kappa$ become independent of each other,
ii) $c$ takes the role of $w_0$ and iii) the spectral index $n_s$ is
more correlated with the number of $e$--folds $N_*$, the curvature
$\hat\kappa$ and the tensor--to--scalar ratio $r$.  This leads to the
conclusion that future CMB and galaxy redshift surveys measuring the
parameters $n_s$, $r$ and $w_0$ with precision should be able to test
a scale invariant model with consistency conditions. The inflationary
observables alone would fix then the values of the model parameters
$c$ and $\kappa$, while the measurement of $w_0$ would provide an
independent test of the consistency relation.

\section{Conclusions}
\label{sec:concl}

Biscalar theories invariant under scale transformations and
volume--preserving diffeomorphisms can accommodate an inflationary
expansion of the Universe followed by a standard hot Big Bang
evolution and a dark-energy dominated era.

The scalar character of the metric determinant under volume-preserving
diffeomorphisms together with the requirement of classical scale
invariance leads to a very specific particle spectrum containing two
graviton polarizations and two scalar degrees of freedom on top of the
standard matter content. A Lagrangian constructed within this
framework contains in general arbitrary functions of the ratio of
these two scalar fields.

In spite of its apparent arbitrariness, the resulting theories turn
out to be predictive. On the one hand, the existence of an effectively
conserved current related to dilatations makes these models
essentially indistinguishable from single--field inflationary
scenarios, from which they ``inherit'' all their virtues. On the other
hand, the symmetries of the Einstein-frame kinetic sector
significantly restrict the inflationary observables. More
specifically, if this target space is maximally symmetric, the
arbitrary functions in the Lagrangian become related in a rather
nontrivial way. As a result, the dynamics is governed by the pole
structure of the Einstein--frame kinetic sector, making the
inflationary predictions universal and almost insensitive to the
details of the potential.

At low energies, the invariance under volume--preserving
diffeomorphisms gives rise to a~\emph{unique} run--away potential for
the dilaton, which can play the role of dynamical dark
energy. Interestingly, the early and late Universe dynamics may become
intertwined in some particular scenarios, leading to non-trivial
consistency relations among the inflationary and dark--energy
observables. The comparison of particular realizations of our paradigm
with present data reveals a strong preference for maximally--symmetric
models with consistency relations over those without
them. Surprisingly, there also is \textit{positive evidence} for the
former class of models over the concordance $\Lambda$CDM model given
the present data sets.

The results of this paper illustrate the strong impact that our
assumptions concerning the early and late Universe dynamics could have
on the interpretation of cosmological data sets. This poses an
interesting question for future CMB observations and galaxy redshift
surveys:

\leavevmode

\noindent\emph{Are inflation and dark energy independent processes in
the expansion history of the Universe or rather two sides of a single
underlying principle?}

\section*{Acknowledgements}

We thank Manuel Trashorras for collaboration in the early stages of
this project and the anonymous referee for useful and constructive
comments. S.C. acknowledges support from CNRS and CNES grants. The
work of G.K.K. is supported by the ERC-AdG-2013 grant 339169
``Selfcompletion.'' M.P. acknowledges support by the Heidelberg
Graduate School of Fundamental Physics as well as the state of
Baden-Württemberg through bwHPC. J.R. acknowledges support from the
Deutsche Forschungsgemeinschaft through the project TRR33 ``The Dark
Universe'' during the first stages of this work.

\newpage
\begin{multicols}{2}
\footnotesize{
\bibliographystyle{utphys}
\bibliography{references}}
\end{multicols}
\end{document}